\def\anon{1}
\def\JASAAuthorName{Bahadır Yüzbaşı}
\def\JASAAffiliation{Department of Econometrics, Faculty of Economics and Administrative Sciences,\\
İnönü University, Malatya, Türkiye}
\def\JASACorrespondingEmail{\href{mailto:bahadir.yuzbasi@inonu.edu.tr}{bahadir.yuzbasi@inonu.edu.tr}}
\def\JASAFundingStatement{No funding was obtained for this work.}
\def\JASACompetingStatement{The author reports no competing interests to declare.}
\def\JASACodeAvailability{The \texttt{HDMaxShrink} R package is available at
\url{https://github.com/byuzbasi/HDMaxShrink}. Study-specific frozen
reproducibility materials are supplied separately for review; the package
repository does not contain the complete study archive.}
\def\JASAGenerativeAIStatement{During manuscript preparation, ChatGPT
(OpenAI) was used for English-language editing and stylistic refinement. The
authors reviewed all content and take full responsibility for the accuracy
and scientific integrity of the manuscript.}
\PassOptionsToPackage{unicode}{hyperref}
\PassOptionsToPackage{hyphens}{url}
\PassOptionsToPackage{dvipsnames,svgnames,x11names}{xcolor}
\documentclass[12pt]{article}

\usepackage{amsmath,amssymb,amsthm,mathtools}
\usepackage{booktabs,longtable,array,multirow}
\usepackage{etoolbox}
\usepackage{graphicx}
\usepackage{xcolor}
\usepackage[font=normalsize,labelfont=bf,labelsep=period,
  justification=raggedright,singlelinecheck=false]{caption}
\usepackage{enumitem}
\usepackage{placeins}
\usepackage[round,authoryear]{natbib}
\usepackage[colorlinks=true,allcolors=blue!55!black]{hyperref}
\usepackage{bookmark}
\IfFileExists{xurl.sty}{\usepackage{xurl}}{}
\usepackage{iftex}
\ifPDFTeX
  \usepackage[T1]{fontenc}
  \usepackage[utf8]{inputenc}
  \usepackage{textcomp}
  \usepackage{lmodern}
\else
  \usepackage{unicode-math}
  \defaultfontfeatures{Scale=MatchLowercase}
  \defaultfontfeatures[\rmfamily]{Ligatures=TeX,Scale=1}
\fi
\IfFileExists{microtype.sty}{\usepackage{microtype}}{}

\makeatletter
\def\HDMSLTEndGlue{\vskip\z@ plus1fil minus1pt\relax}
\patchcmd{\LT@output}
  {\copy\LT@foot\vss}{\copy\LT@foot\HDMSLTEndGlue}
  {}{\PackageError{HDMaxShrink}{First longtable compatibility patch failed}
    {Inspect the installed longtable output routine before compiling.}}
\patchcmd{\LT@output}
  {\copy\LT@foot\vss}{\copy\LT@foot\HDMSLTEndGlue}
  {}{\PackageError{HDMaxShrink}{Second longtable compatibility patch failed}
    {Inspect the installed longtable output routine before compiling.}}
\makeatother

\newtheorem{theorem}{Theorem}
\newtheorem{proposition}[theorem]{Proposition}

\newtheorem{corollary}[theorem]{Corollary}
\newtheorem{assumption}{Assumption}
\theoremstyle{definition}

\newcommand{\R}{\mathbb R}
\newcommand{\E}{\mathbb E}
\newcommand{\Pp}{\mathbb P}
\newcommand{\ind}{\mathbf 1}
\newcommand{\FM}{\mathrm{FM}}
\newcommand{\SM}{\mathrm{SM}}
\newcommand{\PT}{\mathrm{PT}}
\newcommand{\PS}{\mathrm{PS}}

\providecommand{\anon}{0}
\providecommand{\JASAAuthorName}{Author information withheld}
\providecommand{\JASAAffiliation}{Affiliation information withheld}
\providecommand{\JASACorrespondingEmail}{Corresponding-author information withheld}
\providecommand{\JASAFundingStatement}{Funding information is supplied in the identified manuscript.}
\providecommand{\JASACompetingStatement}{Competing-interest information is supplied in the identified manuscript.}
\providecommand{\JASAGenerativeAIStatement}{}

\hypersetup{
  pdftitle={Max-Test-Calibrated Stein Shrinkage with Honest Submodel Selection in Ultra-High-Dimensional Regression},
  pdfkeywords={preliminary-test estimation, conditional Monte Carlo calibration, complementary-pairs stability selection, Ridge regularization, post-selection risk},
  pdfcreator={pdfLaTeX using the official 2025 JASA template}
}
\if1\anon
  \hypersetup{
    pdfauthor={\JASAAuthorName},
    pdfsubject={Statistical methodology preprint with supplementary material}
  }
\else
  \hypersetup{
    pdfauthor={Anonymous},
    pdfsubject={Anonymous submission to JASA Theory and Methods}
  }
\fi

\begin{document}

\def\spacingset#1{\renewcommand{\baselinestretch}{#1}\small\normalsize}
\spacingset{1}

\if1\anon
{
  \bigskip\bigskip\bigskip
  \begin{center}
  \parbox{0.96\textwidth}{%
    \centering\LARGE\bfseries
    Max-Test-Calibrated Stein Shrinkage with\\[0.18em]
    Honest Submodel Selection in\\[0.18em]
    Ultra-High-Dimensional Regression\par
  }

  \bigskip
  {\large \JASAAuthorName\footnotemark}

  \medskip
  \JASAAffiliation
  \end{center}
  \footnotetext{Corresponding author: \JASACorrespondingEmail}
  \medskip
}\fi

\if0\anon
{
  \bigskip\bigskip\bigskip
  \begin{center}
  \parbox{0.96\textwidth}{%
    \centering\LARGE\bfseries
    Max-Test-Calibrated Stein Shrinkage with\\[0.18em]
    Honest Submodel Selection in\\[0.18em]
    Ultra-High-Dimensional Regression\par
  }
  \end{center}
  \medskip
}\fi

\bigskip
\begin{abstract}
Classical preliminary-test and Stein-type estimators interpolate between restricted
and full regression fits, but OLS and chi-squared calibration fail when
$p\gg n$ and the restriction is data-adaptive.  We propose an honest
sample-separated framework built around a common selected-null law.
Independent selection data extend a mandatory core; separate estimation data
pair a complete-design regularized full-model (FM) fit with a fresh
exact-null submodel refit.  A maximum residual-association statistic assesses
all excluded coordinates.  Projected Gaussian draws reproduce its conditional
null distribution under homoskedastic Gaussian errors, yielding a
finite-sample-valid rank test.  The inverse moment of the same design-specific
law replaces $q-2$ and calibrates preliminary-test (PT), Stein-type (S), and
positive-part Stein-type (PS) rules.  We derive exact endpoint-risk formulas,
honest conditional validity, null-law concentration, inverse-moment
consistency, an explicit selection-failure remainder, and endpoint adaptivity
without uniform-dominance claims.  Gaussian experiments with 2,000
replications and up to 30,000 predictors compare Ridge and LASSO FMs, tuned
by frozen-fold cross-validation, under a common submodel, test, and
calibration.  PS provides large near-null coefficient-risk gains and approaches the
relevant full-model risk under strong departures.  A separate
CPSS-LASSO/CPSS-MCP audit evaluates data-adaptive selection, while a
split-sample DepMap study with 19,152 predictors illustrates prediction gains
and sensitivity to error assumptions.  The \texttt{HDMaxShrink} R package
implements the procedure.
\end{abstract}

complementary-pairs stability selection; Ridge regularization;
post-selection risk.
\vfill

\newpage
\spacingset{1} 

\section{Introduction}
\label{sec:introduction}

Preliminary-test and Stein-type estimators address a basic model-uncertainty
tradeoff: a credible restriction can sharply reduce risk, whereas an
unrestricted fit is safer when that restriction fails.  Classical procedures
interpolate between these endpoints with a Wald, likelihood-ratio, or
$F$ statistic and a dimension-dependent constant motivated by the
James--Stein principle \citep{jamesStein1961}.  This architecture does not
transfer directly to ultra-high-dimensional regression.  For $p>n$ the full
Gram matrix is singular, for $q=p-p_1\gg n$ a fixed-dimensional chi-squared
calibration is generally unavailable, and a submodel learned from the same
responses used for testing destroys the usual fixed-restriction argument.

Several literatures supply relevant components.  LASSO, stability selection,
and complementary-pairs stability selection (CPSS) provide sparse fitting and
selection-frequency control
\citep{tibshirani1996,meinshausenBuhlmann2010,shahSamworth2013}; sample
separation underlies screen-and-clean and honest estimation
\citep{wassermanRoeder2009,atheyImbens2016}; and debiased scores and multiplier
methods enable simultaneous high-dimensional inference
\citep{zhangZhang2014,vanDeGeerEtAl2014,javanmardMontanari2014,ningLiu2017,zhangCheng2017,dezeureBuhlmannZhang2017}.
Post-selection refitting and high-dimensional Ridge risk are also well
understood in their respective settings
\citep{belloniChernozhukov2013,dobribanWager2018}.  Separately,
preliminary-test and Stein shrinkage have been combined with penalized and
high-dimensional regression
\citep{salehRaheem2015,norouziradArashi2018,gaoAhmedFeng2017,ahmedYuzbasi2017,yuzbasiArashiAhmed2020}.
A unified treatment of full-model/submodel shrinkage, penalized selection,
and weak-signal integration, including high-dimensional regression, is given
by \citet{ahmedAhmedYuzbasi2023}.  What is missing is a single construction in
which a data-adaptive restriction,
a valid $q\gg n$ diagnostic, and the shrinkage calibration share one coherent
selected-null law.

Our central object is the \emph{selected coordinate null}.  An independent
selection sample uses mandatory-core CPSS to produce a small index set
$\widehat A$.  A separate estimation sample defines two endpoints: an
ordinary Ridge full-model (FM) fit using the complete design $X$, computed in
its $n$-dimensional dual form, and a submodel (SM) singular-value refit using
$X_{\widehat A}$ with all excluded
coefficients set exactly to zero.  Thus the SM is a genuine restricted
estimator, not an oracle and not a copy of the FM coefficients.  Neither
endpoint invokes full-model OLS or an inverse of $X^{\mathsf T}X$.

To assess the restriction, we project the response and each excluded
predictor away from the selected core and take the largest absolute
standardized residual association.  Conditional on the design and selection
sample, applying this same maximum statistic to projected Gaussian draws
reproduces its null law under homoskedastic Gaussian errors.  The resulting
rank Monte Carlo test is therefore finite-sample valid even when $q\gg n$.
It targets sparse departures and is not presented as a high-dimensional Wald
or partial-$F$ test; a multiplier version is used only as an asymptotic
heteroskedasticity sensitivity analysis.

The same conditional law also supplies the Stein calibration.  If $T_0^*$ is
a null draw, define

\[
\kappa_A(X)=\{\E[(T_0^*)^{-2}\mid X,A]\}^{-1}.
\]

For a classical chi-squared statistic, the corresponding inverse moment
reduces to $q-2$.  Here the correlated maximum-statistic law replaces that
constant for the preliminary-test (PT), Stein-type (S), and positive-part
Stein-type (PS) estimators.  This common-law construction---one simulated
null for both
testing and interpolation---is the main methodological novelty.  It centers
the untruncated Stein-form weight but does not imply uniform James--Stein
dominance.

The article makes four contributions.

\begin{enumerate}[leftmargin=*,itemsep=3pt]
\item It gives an honest mandatory-core selection architecture and two
well-defined $p\gg n$ endpoints: a complete-design Ridge FM and a fresh
exact-null SVD SM.
\item It derives a conditionally exact Gaussian maximum test for the selected
coordinate null and uses its design-specific inverse moment to calibrate the
PT, S, and PS estimators.
\item It establishes endpoint risk formulas, honest conditional validity,
pointwise interpolation and endpoint-mapping results, inverse-moment
consistency, and an explicit remainder for selection failure, while stating
clearly where uniform dominance is unavailable.
\item It evaluates Ridge and LASSO FMs under a common restriction and
calibration, with frozen-fold cross-validation; audits CPSS--LASSO and
CPSS--MCP on independent selection and analysis samples; and illustrates the method in a split-sample DepMap
analysis.  The accompanying \texttt{HDMaxShrink} package provides a
reproducible implementation.
\end{enumerate}

The construction is therefore distinct from the generalized-Ridge framework
of \citet{yuzbasiArashiAhmed2020}: Ridge is only one FM endpoint here, whereas
the defining elements are an honestly selected exact-null SM, a conditionally
simulated maximum diagnostic, and a null-law inverse moment in place of
$q-2$.  The contribution is their joint geometry and theory, not the
individual invention of its component methods.

Section~\ref{sec:methodology} defines the selection rule, endpoints, test, and
estimators; Section~\ref{sec:theory} gives the theoretical results; and
Section~\ref{sec:computation} describes computation.
Sections~\ref{sec:simulation-design}--\ref{sec:results} present the numerical and
DepMap studies, and Section~\ref{sec:discussion} discusses scope and
limitations.  Proofs and extended numerical diagnostics are provided in
Supplementary Appendices A--D.

\section{Methodology}
\label{sec:methodology}

\subsection{Model, honest splitting, and notation}

Let

\[
y=X\beta^0+\varepsilon,
\qquad X\in\R^{n\times p},
\qquad p\gg n,
\]

where $\beta^0$ is sparse and, for the primary finite-sample theory,
$\varepsilon\mid X\sim N_n(0,\sigma^2I_n)$.  The selection sample
$\mathcal D_S=(X_S,y_S)$ is independent of the estimation/inference sample
$\mathcal D_E=(X_E,y_E)$.  The core and data-dependent penalty are measurable
with respect to $\mathcal D_S$ and frozen before endpoint estimation and
testing on $\mathcal D_E$.  This \emph{honest} split follows screening and
cleaning \citep{wassermanRoeder2009} and honest estimation
\citep{atheyImbens2016}; CPSS alone does not provide honesty, support recovery,
an oracle property, or uniform confidence-interval coverage.

For a generic sample of size $n$, define

\[
H_n=I_n-n^{-1}{\bf1}{\bf1}^{\mathsf T},\qquad
X_c=H_nX,\qquad y_c=H_ny.
\]

Predictors and response are root-mean-square standardized within each
training sample.  Write

\[
s_{xj}^2=n^{-1}\|X_{c,j}\|_2^2,\qquad
D_x=\operatorname{diag}(s_{x1},\ldots,s_{xp}),\qquad
s_y^2=n^{-1}\|y_c\|_2^2,
\]
\[
X_s=X_cD_x^{-1},\qquad y_s=y_c/s_y.
\]

All empirical scales must be positive.  Computation uses zero standardized
intercept; $\widehat\theta$ is returned to the original scale as
$\widehat\beta=s_yD_x^{-1}\widehat\theta$, with

\[
\widehat\beta_0=\bar y-\bar x^{\mathsf T}\widehat\beta
\]

This reconstructs the intercept.  No full-model OLS estimator is defined or
used.

\subsection{Mandatory-core conditional CPSS}

A mandatory core $A_0\subset\{1,\ldots,p\}$, fixed without using $y_E$,
contains scientifically required covariates, design factors, or a frozen base
model.  The optional family is

\[
\mathcal C=\{1,\ldots,p\}\setminus A_0.
\]

Mandatory predictors are always retained and lie outside the CPSS
false-selection family.  For each complementary half-sample $I$ from
$\mathcal D_S$, a checked economy-SVD basis $Q_{0,I}$ of the centered
mandatory design gives the Frisch--Waugh--Lovell residuals

\[
\widetilde y_I=(I-Q_{0,I}Q_{0,I}^{\mathsf T})H_Iy_I,
\qquad
\widetilde X_{I,\mathcal C}
=(I-Q_{0,I}Q_{0,I}^{\mathsf T})H_IX_{I,\mathcal C}.
\]

A base selector $g\in\{L,M\}$ (LASSO or MCP) is fitted only to these
residualized optional variables under a frozen half-sample selection budget.
With paired selection frequency $\widehat\Pi_{j,g}$,

\[
\widehat E_g
=\{j\in\mathcal C:\widehat\Pi_{j,g}\ge\pi_{\rm thr}\},
\qquad
\widehat A_g=A_0\cup\widehat E_g.
\]

The primary selector is CPSS--LASSO with $\pi_{\rm thr}=0.60$; CPSS--MCP is a
sensitivity analysis because its selected path can leave the locally convex
region.  Threshold $0.25$ is exploratory and has no advertised error-control
guarantee.  An empty extension gives $\widehat A_g=A_0$ without forcing an
optional variable.  Rank, degeneracy, duplication, convergence, and selected-
size diagnostics are recorded for every half-sample.

The familiar stability-selection expression

\[
\frac{q_\Lambda^2}
{(2\pi_{\rm thr}-1)|\mathcal C_{\rm eligible}|}
\]

is only an assumption-dependent expected false-selection bound for the
optional family when $\pi_{\rm thr}>1/2$ \citep{meinshausenBuhlmann2010}; the
sharper CPSS bounds require the target-set and shape conditions of
\citet{shahSamworth2013}.

\subsection{Ridge FM endpoint}

The Ridge penalty $\widehat\lambda_R>0$ is selected by frozen-fold
cross-validation on $\mathcal D_S$ and held fixed on $\mathcal D_E$.  Put
$a=n\widehat\lambda_R$ and $S=X_c^{\mathsf T}X_c$.  The standardized FM is

\[
\widehat\theta^{\FM}
=X_s^{\mathsf T}(X_sX_s^{\mathsf T}+aI_n)^{-1}y_s.
\]

The implemented FM endpoint on the original coefficient scale is therefore

\begin{align}
\widehat\beta^{\FM}
&=s_yD_x^{-1}X_s^{\mathsf T}
(X_sX_s^{\mathsf T}+aI_n)^{-1}y_s\nonumber\\
&=(S+aD_x^2)^{-1}X_c^{\mathsf T}y_c.
\label{eq:ridge-fm}
\end{align}

The response scale cancels, whereas predictor scaling remains through
$aD_x^2$ (equal to $aI_p$ only when all RMS scales are one).  This is isotropic
Ridge on standardized predictors, computed through the $n\times n$ dual
system.  Every predictor enters one fit; there is no $X_A$/$X_B$ penalty split,
screening, or full-model OLS inverse.

\subsection{Exact-null SVD submodel endpoint}

Fix one independently selected core $A=\widehat A_g$, set $B=A^c$, and write
$p_1=|A|<n-1$ and $q=|B|$.  The selected restriction is

\begin{equation}
H_0(A):\beta_B^0=0_q.
\label{eq:selected-null}
\end{equation}

The SM estimator refits under the null in Equation~\eqref{eq:selected-null}:

\begin{equation}
\widehat\beta_A^{\SM}=X_{c,A}^{+}y_c,
\qquad
\widehat\beta_B^{\SM}=0_q,
\label{eq:sm}
\end{equation}

In Equation~\eqref{eq:sm}, $X_{c,A}^{+}$ is computed by checked economy SVD.  For a checked
full-column-rank core, standardize--backscale and original-scale refits agree.
The retained SM coefficients are fitted anew, not copied from Ridge; zeroing
the complement of an FM fit would not be a restricted refit.

\subsection{Maximum partial-\texorpdfstring{$t$}{t} restriction diagnostic}

Let $Q_A$ be an orthonormal basis for $\operatorname{col}(X_{c,A})$ and define

\[
M_A=H_n-Q_AQ_A^{\mathsf T},\qquad
r_0=M_Ay,\qquad z_j=M_AX_j\quad(j\in B).
\]

After removing only numerically zero $\|z_j\|_2$ directions, put

\[
d=n-\operatorname{rank}(X_{c,A})-1,
\qquad
\widehat\sigma_0=\|r_0\|_2/\sqrt d
\]

and require $d>1$ for the partial-$t$ transformation; the SM refit itself only
requires $p_1<n-1$.  Calculate

\[
T_{\rm score}
=\max_{j\in B}
\frac{|z_j^{\mathsf T}r_0|}
{\|z_j\|_2\widehat\sigma_0}.
\]

The reported monotone partial-$t$ transformation is

\begin{equation}
T_{\max}
=\left\{\frac{(d-1)T_{\rm score}^2}
{d-T_{\rm score}^2}\right\}^{1/2}.
\label{eq:tmax}
\end{equation}

For $b=1,\ldots,B_{MC}$, apply the identical map to
$g_b\sim N_n(0,I_n)$ to obtain $T_{\max}^{*(b)}$.  The rank p-value and
decision are

\begin{equation}
\widehat p
=\frac{1+\sum_{b=1}^{B_{MC}}
\ind\{T_{\max}^{*(b)}\ge T_{\max}\}}
{B_{MC}+1},
\qquad
R=\ind\{\widehat p\le\alpha\}.
\label{eq:mc-p}
\end{equation}

This sparse-departure diagnostic is not Wald; it uses neither a $\chi_q^2$ nor
a joint partial-$F$ law.

\subsection{Inverse-null-moment calibration and estimator family}

The same conditional draws define

\begin{equation}
\widehat\kappa
=\left\{B_{MC}^{-1}\sum_{b=1}^{B_{MC}}
(T_{\max}^{*(b)})^{-2}\right\}^{-1}.
\label{eq:kappa}
\end{equation}

Let

\[
D=\widehat\beta^{\FM}-\widehat\beta^{\SM},
\qquad
W=1-\frac{\widehat\kappa}{T_{\max}^2\vee\epsilon},
\qquad W_+=\max(0,W),
\]

with a fixed numerical guard $\epsilon=10^{-10}$.  Equations~\eqref{eq:pt}
and~\eqref{eq:s} use binary and untruncated weights, respectively:

\begin{align}
\widehat\beta^{\PT}
&=\widehat\beta^{\SM}+RD,
\label{eq:pt}\\
\widehat\beta^{S}
&=\widehat\beta^{\SM}+WD,
\label{eq:s}\\
\widehat\beta^{\PS}
&=\widehat\beta^{\SM}+W_+D.
\label{eq:ps}
\end{align}

Equations~\eqref{eq:pt}--\eqref{eq:ps} define the PT, S, and PS estimators,
respectively.  PS is the primary continuous estimator; its weight uses the calibrated
statistic magnitude, not only rejection.  The inverse moment in
\eqref{eq:kappa} is not $q-2$ and does not imply classical Stein dominance.

\subsection{Robust test-only sensitivity}

For non-Gaussian real data, centered score contributions
$\psi_{ij}=z_{ij}r_{0i}$ are coordinatewise studentized and their maximum is
approximated with Gaussian multipliers
\citep{chernozhukovChetverikovKato2013}.  This heteroskedasticity-robust
asymptotic sensitivity diagnostic neither replaces the Gaussian draws in
\eqref{eq:kappa} nor changes PT/S/PS weights.

\section{Theory}
\label{sec:theory}

Unless stated otherwise, expectations condition on the selection sample, so
$A$ is fixed before analysis of the estimation sample.  Risks use original-
scale coefficients; $X_A$ and $X_B$ abbreviate $X_{c,A}$ and $X_{c,B}$, and
$D_x$ is the positive RMS scale matrix from Section~\ref{sec:methodology}.
Detailed derivations and all proofs are given in the Supplementary Material.

\begin{assumption}[Primary Gaussian model]
\label{ass:gaussian}
Conditional on $X$, the estimation-sample errors are
$N_n(0,\sigma^2I_n)$.  The Ridge penalty is positive and measurable with
respect to the independent selection sample.  The selected centered design
$X_A$ has rank $p_1<n-1$, and at least one excluded predictor has a nonzero
residualized direction under $M_A$.
\end{assumption}

\subsection{Exact risk of the two endpoints}

The endpoint decompositions quantify the cost of a false selected null.

\begin{proposition}[Conditional SM risk]
\label{prop:sm-risk}
Under Assumption~\ref{ass:gaussian},
\begin{align}
R_{\SM}(X,A)
&=\E\!\left(
\|\widehat\beta^{\SM}-\beta^0\|_2^2\mid X,A
\right)\nonumber\\
&=\|\beta_B^0\|_2^2
+\|X_A^+X_B\beta_B^0\|_2^2
+\sigma^2\|X_A^+\|_F^2.
\label{eq:sm-risk}
\end{align}
\end{proposition}

The three terms in \eqref{eq:sm-risk} are exclusion bias, aliasing into the
retained coordinates, and refit variance; no full-model inverse appears.

\begin{corollary}[Gaussian-iid SM benchmark]
\label{cor:sm-iid}
Suppose the uncentered rows of $X$ are iid $N_p(0,I_p)$, independently of the
errors, and $n>p_1+2$.  Then
\begin{equation}
R_{\SM}
=\|\beta_B^0\|_2^2
+(\sigma^2+\|\beta_B^0\|_2^2)
\frac{p_1}{n-p_1-2}.
\label{eq:sm-risk-iid}
\end{equation}
\end{corollary}

At $\beta_B^0=0$, \eqref{eq:sm-risk-iid} reduces to
$\sigma^2p_1/(n-p_1-2)$; a norm-$\Delta$ departure adds
$\Delta^2\{1+p_1/(n-p_1-2)\}$.

For FM, put $S=X_c^{\mathsf T}X_c$, $a=n\widehat\lambda_R$, and
$K_a=S+aD_x^2$; the conditioning fixes $a>0$ and $D_x$.

\begin{proposition}[Conditional Ridge risk]
\label{prop:fm-risk}
For every $a>0$, the Ridge estimator in \eqref{eq:ridge-fm} equals
$K_a^{-1}X_c^{\mathsf T}y_c$, and its original-scale coefficient risk is
\begin{equation}
R_{\FM}(X)
=a^2\|K_a^{-1}D_x^2\beta^0\|_2^2
+\sigma^2\operatorname{tr}(K_a^{-1}SK_a^{-1}).
\label{eq:fm-risk}
\end{equation}
\end{proposition}

The first term includes the unavoidable null-space bias under full-vector
loss when $p>n$; deterministic equivalents require an additional signal model
\citep{dobribanWager2018}.  This term also changes the classical large-
departure RPE pattern.

\begin{proposition}[Large-departure Ridge/SM RPE plateau]
\label{prop:large-delta-plateau}
Fix $X$, its RMS scale matrix $D_x$, $A$, and $a=n\widehat\lambda_R>0$.  Let
$\beta^0(\Delta)=\beta^{(0)}+\Delta v$, where $v_A=0$ and
$v_B\ne0$.  Then, as $|\Delta|\to\infty$,
\begin{equation}
\frac{R_{\FM}(X;\Delta)}{R_{\SM}(X,A;\Delta)}
\longrightarrow
\frac{a^2\|K_a^{-1}D_x^2v\|_2^2}
{\|v_B\|_2^2+\|X_A^+X_Bv_B\|_2^2}>0.
\label{eq:large-delta-plateau}
\end{equation}
If $a\downarrow0$ after the design is fixed, then
\[
aK_a^{-1}D_x^2v
\longrightarrow
D_x^{-1}P_{\operatorname{null}(X_s)}D_xv.
\]
Here $P_{\operatorname{null}(X_s)}$ is the Euclidean orthogonal projection
in standardized coordinates.  Its conjugate by $D_x$ is generally oblique
in the original coefficient metric.  A nonzero projected departure
therefore retains a null-space contribution even as the penalty vanishes.
\end{proposition}

Thus Ridge/SM RPE need not vanish as it can under full-rank OLS.  LASSO can
behave differently under separate design, sparsity, tuning, and recovery
conditions, without changing the primary Ridge FM endpoint.

\subsection{Conditional Monte Carlo validity}

Let $\mathcal T_A(v;X)$ denote the complete map used to calculate
$T_{\max}$ from a generic vector $v$: centering, projection by $M_A$,
self-normalization, maximization over the checked residualized directions,
and the transformation in \eqref{eq:tmax}.

\begin{theorem}[Finite-sample conditional validity]
\label{thm:mc-validity}
Under Assumption~\ref{ass:gaussian} and the selected null
$H_0(A):\beta_B^0=0_q$, suppose additionally that
$d=n-p_1-1>1$.  The statistic
$T_{\max}=\mathcal T_A(\varepsilon;X)$ and the simulated statistics
$T_{\max}^{*(b)}=\mathcal T_A(g_b;X)$ are exchangeable up to their irrelevant
common scale.  Therefore the p-value in \eqref{eq:mc-p} satisfies
\begin{equation}
\Pp(\widehat p\le\alpha\mid\mathcal D_S,X,H_0(A))\le\alpha.
\label{eq:conditional-size}
\end{equation}
The result permits $q>n$ and arbitrary dependence among excluded design
columns.
\end{theorem}

\begin{corollary}[Transfer through honest selection]
\label{cor:honest-transfer}
For any selection rule measurable with respect to the independent selection
sample, provided that the checked test-domain conditions of
Theorem~\ref{thm:mc-validity} hold almost surely,
\begin{equation}
\Pp\{\widehat p\le\alpha,\ H_0(A)\text{ is true}\}
\le\alpha\Pp\{H_0(A)\text{ is true}\}.
\label{eq:selection-transfer}
\end{equation}
\end{corollary}

Equation~\eqref{eq:selection-transfer} controls size only for true selected nulls;
omitting an active variable affects power and risk instead.  Reusing selection
outcomes for testing is not covered by Theorem~\ref{thm:mc-validity}.

\subsection{Null geometry and Gaussian-width calibration}

Finite-sample validity needs no extreme-value approximation.  To compare the
null maximum with its inverse moment, let $L_n\in\R^{n\times d_n}$ span
$\operatorname{range}(M_{A_n})$ with orthonormal columns and, for each checked
nonzero residualized direction, put
\[
u_{nj}=\frac{L_n^{\mathsf T}z_{nj}}{\|z_{nj}\|_2}
\in\mathbb S^{d_n-1},
\qquad j=1,\ldots,q_n.
\]
Here $q_n$ counts eligible checked directions and may be below the nominal
complement size.

\begin{proposition}[Exact residual-sphere representation]
\label{prop:sphere-representation}
Under $H_0(A_n)$ and Assumption~\ref{ass:gaussian}, let
$G_n=L_n^{\mathsf T}\varepsilon/\sigma\sim N_{d_n}(0,I_{d_n})$ and define
\[
Z_n=\max_{1\le j\le q_n}|u_{nj}^{\mathsf T}G_n|,
\qquad R_n=\|G_n\|_2.
\]
Then, almost surely,
\begin{equation}
T_{\max,n}^2
=\frac{(d_n-1)Z_n^2}{R_n^2-Z_n^2}.
\label{eq:sphere-representation}
\end{equation}
Equation~\eqref{eq:sphere-representation} also represents each conditional Gaussian calibration draw.
\end{proposition}

Write $a_n=\E_G Z_n$ for fixed selected design.  This Gaussian width of
$\{\pm u_{n1},\ldots,\pm u_{nq_n}\}$ incorporates dependence and need not
equal $\sqrt{2\log q_n}$.

\begin{assumption}[Gaussian-width and lower-tail regularity]
\label{ass:gaussian-width}
Along the fixed selected-design sequence,
\begin{equation}
a_n\longrightarrow\infty,
\qquad
\frac{a_n^2}{d_n}\longrightarrow0.
\label{eq:width-rate}
\end{equation}
In addition, for some constants $C<\infty$ and $\gamma>8$, uniformly in $n$,
\begin{equation}
\Pp_G(Z_n\le t a_n\mid X,A_n)\le C t^\gamma,
\qquad 0<t\le1.
\label{eq:width-small-ball}
\end{equation}
The number of calibration draws $B_{MC,n}$ tends to infinity independently
of the observed Gaussian error.
\end{assumption}

Equation~\eqref{eq:width-rate} sets the growth regime;
Equation~\eqref{eq:width-small-ball} rules out excessive lower-tail mass;
$\gamma>8$ supplies the inverse fourth moments used by empirical calibration.

\begin{theorem}[Conditional null concentration]
\label{thm:null-width-concentration}
Under $H_0(A_n)$ and Assumptions~\ref{ass:gaussian} and
\ref{ass:gaussian-width},
\begin{equation}
\frac{T_{\max,n}^2}{a_n^2}\overset p\longrightarrow1,
\qquad
\frac{\kappa_{A_n}(X)}{a_n^2}\longrightarrow1,
\qquad
\frac{\widehat\kappa_n}{a_n^2}\overset p\longrightarrow1.
\label{eq:null-width-concentration}
\end{equation}
Combining the limits in Equation~\eqref{eq:null-width-concentration} gives
\begin{equation}
\frac{T_{\max,n}^2}{\widehat\kappa_n}
\overset p\longrightarrow1,
\qquad
\frac{T_{\max,n}^2\vee\epsilon}{\widehat\kappa_n}
\overset p\longrightarrow1.
\label{eq:null-calibrated-ratio}
\end{equation}
All limits are conditional on the selected-design sequence.  The fixed guard
in the second ratio vanishes relative to $\widehat\kappa_n$ because
$a_n\to\infty$.
\end{theorem}

The result covers dependent designs satisfying the width conditions; the
$2\log q$ scale additionally requires sufficient random-design dispersion.

\begin{corollary}[Gaussian-iid $2\log q$ scale]
\label{cor:iid-max-scale}
Under $H_0(A_n)$ and Assumption~\ref{ass:gaussian}, suppose the uncentered
estimation-design rows are iid $N_p(0,I_p)$, independently of the honest
selection sample, and let $d_n=n-p_{1,n}-1$.  Let $B_{MC,n}\to\infty$,
with calibration draws independent of the observed errors.  If
\begin{equation}
d_n\to\infty,\qquad q_n/d_n\to\infty,\qquad
\frac{\{\log(2q_n)\}^3}{d_n}\to0,
\label{eq:iid-max-rate}
\end{equation}
then the width and small-ball conditions of
Assumption~\ref{ass:gaussian-width} hold on design events with probability
tending to one, with constants independent of sufficiently large $n$
(for any fixed $\gamma>8$), and
\begin{equation}
\frac{a_n^2}{2\log(2q_n)}\longrightarrow1.
\label{eq:iid-width}
\end{equation}
Equation~\eqref{eq:iid-width} yields, jointly over design, error, and calibration draws,
\begin{equation}
\frac{T_{\max,n}^2}{2\log(2q_n)}\overset p\longrightarrow1,
\qquad
\frac{\widehat\kappa_n}{2\log(2q_n)}
\overset p\longrightarrow1.
\label{eq:iid-tmax-kappa}
\end{equation}
Since $\log(2q_n)/\log q_n\to1$, either logarithm gives the same first-order
scale.  Condition~\eqref{eq:iid-max-rate} is sufficient, not asserted to be
necessary.  The conclusion does not extend to arbitrary dependent columns;
sharp lower-tail control for dependent Gaussian maxima is precisely what
prevents nominal $q_n$ from replacing effective multiplicity without a design
condition \citep{lopesYao2022}.
The supplementary proof makes the small-ball step explicit using Gaussian
comparison, concentration, and singular-value bounds
\citep[Lemma~5.33, Proposition~5.34, and Corollary~5.35]{vershynin2012}.
\end{corollary}

\subsection{What the inverse moment guarantees}

Let $T_0^*$ have the conditional null law and suppose its inverse moment is
finite.  Define

\[
\kappa_A(X)=\{\E[(T_0^*)^{-2}\mid X,A]\}^{-1}.
\]

Then

\begin{equation}
\E\left\{1-\frac{\kappa_A(X)}{(T_0^*)^2}
\mathrel{\Big|}X,A\right\}=0.
\label{eq:null-centering}
\end{equation}

Equation~\eqref{eq:null-centering} centers the ideal weight; it is not an unbiased-risk identity.

\begin{proposition}[Monte Carlo calibration consistency]
\label{prop:kappa}
For a fixed checked design, suppose
$\E\{(T_0^*)^{-4}\mid X,A\}<\infty$.  As $B_{MC}\to\infty$,
\begin{equation}
\widehat\kappa\longrightarrow\kappa_A(X)
\qquad\text{almost surely conditionally on }(X,A).
\label{eq:kappa-consistency}
\end{equation}
\end{proposition}

The inverse-moment condition excludes excessive small-ball mass from too few
distinct residual directions; implementation diagnostics monitor this issue.

\subsection{Interpolation geometry and endpoint adaptivity}

For $w\in[0,1]$, define
$\widehat\beta(w)=\widehat\beta^{\SM}+wD$.

\begin{proposition}[Pointwise loss identity]
\label{prop:loss-identity}
For every realized data set and every $\beta^0$,
\begin{align}
\|\widehat\beta(w)-\beta^0\|_2^2
&=(1-w)\|\widehat\beta^{\SM}-\beta^0\|_2^2
+w\|\widehat\beta^{\FM}-\beta^0\|_2^2\nonumber\\
&\quad-w(1-w)\|\widehat\beta^{\FM}
-\widehat\beta^{\SM}\|_2^2.
\label{eq:loss-identity}
\end{align}
\end{proposition}

By Equation~\eqref{eq:loss-identity}, PS lies below the convex average, not necessarily below
both realized endpoint losses.

Under an alternative, define the residual noncentrality and its largest
checked projection by
\[
h_n=\sigma^{-1}L_n^{\mathsf T}M_{A_n}X_{B_n}\beta^0_{B_n},
\qquad
\lambda_n=\max_{1\le j\le q_n}|u_{nj}^{\mathsf T}h_n|.
\]

\begin{theorem}[Separated-alternative concentration]
\label{thm:separated-alternative}
Suppose the selected-design geometry satisfies
Assumption~\ref{ass:gaussian-width}, so that the Gaussian calibration obeys
$\widehat\kappa_n/a_n^2\overset p\longrightarrow1$.  If
\begin{equation}
\frac{d_n\lambda_n^2}
{a_n^2\{d_n+\|h_n\|_2^2\}}
\longrightarrow\infty,
\label{eq:separated-alternative}
\end{equation}
then
\begin{equation}
\frac{T_{\max,n}^2}{\widehat\kappa_n}
\overset p\longrightarrow\infty.
\label{eq:alternative-calibrated-ratio}
\end{equation}
A simpler sufficient condition is
$\|h_n\|_2^2=O(d_n)$ and $\lambda_n/a_n\to\infty$.
\end{theorem}

Separation is in residualized score geometry: a large but aliased coefficient
need not make $\lambda_n$ large.

\begin{theorem}[PS endpoint mapping]
\label{thm:endpoint-mapping}
Consider a sequence of selected designs and let
$0<\widehat\kappa<\infty$ almost surely.  With the implemented fixed guard
$\epsilon=10^{-10}$, define
\[
V_n=\frac{T_{\max}^2\vee\epsilon}{\widehat\kappa},\qquad
w_{\PS}=(1-V_n^{-1})_+.
\]
If
\begin{equation}
V_n\overset p\longrightarrow1,
\label{eq:null-ratio}
\end{equation}
then $w_{\PS}\overset p\longrightarrow0$.  If instead
\begin{equation}
V_n\overset p\longrightarrow\infty,
\label{eq:alternative-ratio}
\end{equation}
then $w_{\PS}\overset p\longrightarrow1$.  Consequently, for any
deterministic sequence $r_n>0$ with $\|D\|_2=O_p(r_n)$,
\begin{align*}
\widehat\beta^{\PS}-\widehat\beta^{\SM}&=o_p(r_n)
&&\text{under \eqref{eq:null-ratio}},\\
\widehat\beta^{\PS}-\widehat\beta^{\FM}&=o_p(r_n)
&&\text{under \eqref{eq:alternative-ratio}}.
\end{align*}
\end{theorem}

For $U_n=T_{\max}^2/\widehat\kappa$, $U_n\to1$ additionally requires
$\epsilon/\widehat\kappa\to0$ to imply \eqref{eq:null-ratio}; by contrast,
$U_n\to\infty$ alone implies \eqref{eq:alternative-ratio}.  Theorems
\ref{thm:null-width-concentration} and \ref{thm:separated-alternative} provide
the two premises.  Multiplicity remains geometric: perfectly correlated
columns preclude an automatic $2\log q$ law.

\begin{corollary}[Endpoint risk transfer]
\label{cor:endpoint-risk-transfer}
For $E\in\{\SM,\FM\}$, write
\[
e_{E,n}=\widehat\beta_n^E-\beta_n^0,\qquad
R_{E,n}=\E\|e_{E,n}\|_2^2,
\]
where the expectation includes the estimation error and calibration draws
under the same fixed selected-design conditioning as above, and suppose the
relevant $R_{E,n}$ is positive.

Under $H_0(A_n)$ and the assumptions of
Theorem~\ref{thm:null-width-concentration}, if
\begin{equation}
\left\{
\frac{(\|e_{\SM,n}\|_2+\|D_n\|_2)^2}{R_{\SM,n}}
\right\}_{n\ge1}
\quad\text{is uniformly integrable},
\label{eq:sm-risk-ui}
\end{equation}
then
\[
\frac{R_{\PS,n}}{R_{\SM,n}}\longrightarrow1.
\]
Under the assumptions of Theorem~\ref{thm:separated-alternative}, if
\begin{equation}
\left\{
\frac{(\|e_{\FM,n}\|_2+\|D_n\|_2)^2}{R_{\FM,n}}
\right\}_{n\ge1}
\quad\text{is uniformly integrable},
\label{eq:fm-risk-ui}
\end{equation}
then
\[
\frac{R_{\PS,n}}{R_{\FM,n}}\longrightarrow1.
\]
\end{corollary}

These are endpoint-equivalence statements.  They neither order the finite-
sample risks nor imply that PS dominates both endpoints between the null and
separated-alternative regimes.

At fixed $\alpha$, valid PT rejection need not vanish, so null equivalence to
SM is not automatic.  A sufficient, not necessary, route is
$\alpha_n\to0$ with control of $D$; under separated alternatives and
consistent rejection, PT maps to FM.

\subsection{Selection failures and the scope of CPSS theory}

Let $S_0=\operatorname{supp}(\beta^0)$ and define

\[
\mathcal E_n=\{S_0\subseteq\widehat A,
\ \operatorname{rank}(X_{\widehat A})=|\widehat A|,
\ |\widehat A|<n-1\}.
\]

For any coefficient estimate $b\in\R^p$, write
$L_2(b,\beta^0)=\|b-\beta^0\|_2^2$ for its squared Euclidean coefficient
loss.  This is the coefficient-risk loss used throughout the numerical study.

\begin{proposition}[Selection-failure remainder]
\label{prop:selection-remainder}
For any estimator $\widehat\beta^\star$ in the proposed family,
\begin{equation}
\E\{L_2(\widehat\beta^\star,\beta^0)
\ind(\mathcal E_n^c)\}
\le
\{\E L_2(\widehat\beta^\star,\beta^0)^2\}^{1/2}
\Pp(\mathcal E_n^c)^{1/2}.
\label{eq:selection-remainder}
\end{equation}
\end{proposition}

Equation~\eqref{eq:selection-remainder} bounds selection-failure risk;
oracle-like approximation also needs sure screening and controlled fourth
moments.  Fixed-threshold CPSS does not ensure support
consistency; its false-selection bounds require their stated conditions,
apply only to the optional family, and do not control false negatives
\citep{meinshausenBuhlmann2010,shahSamworth2013}.

\subsection{Non-dominance statement}

The results above establish exact endpoint risks, valid testing, calibration
consistency, interpolation geometry, and endpoint mapping.  They do not
establish finite-sample or uniform dominance of S or PS over FM.  Such a
claim would require a Stein identity tailored to the joint nonlinear map from
the response to the selected core, maximum statistic, estimated inverse
moment, and Ridge/SVD endpoints.  The simulations therefore report where
each method gains and loses rather than treating dominance as a premise.

\section{Computation}
\label{sec:computation}

The implementation is designed for $p\gg n$ and never forms or inverts a
full $p\times p$ Gram matrix.

\subsection{Endpoint linear algebra}

Rank-sensitive operations use economy SVD with an explicit relative
singular-value tolerance.  This covers mandatory-core residualization, the
exact-null SM refit, and construction of $M_A$.  Rank loss in a mandatory
core is reported as a failure rather than repaired by silently deleting a
required predictor; optional predictors annihilated by residualization are
recorded as ineligible.

For Ridge, one eigendecomposition of $X_sX_s^{\mathsf T}$ evaluates the
positive penalty grid.  A fit at fixed penalty has leading cost
$O(n^2p+n^3)$ and requires only the design and an $n\times n$ system.  Each
solution is checked through the standardized first-order residual

\[
\left\|n^{-1}X_s^{\mathsf T}
(X_s\widehat\theta^{\FM}-y_s)
+\lambda_R\widehat\theta^{\FM}\right\|_\infty,
\qquad
\widehat\theta^{\FM}=D_x\widehat\beta^{\FM}/s_y.
\]

\subsection{Conditional simulation and CPSS}

The residualized directions $z_j/\|z_j\|_2$ are computed once per test.
Gaussian draws are then processed in blocks, giving $O(B_{MC}nq)$ arithmetic
without storing a $B_{MC}\times q$ matrix.  The observation and all draws use
the same checked directions, and the statistic, critical value, rank
p-value, inverse-moment calibration, and Monte Carlo uncertainty are retained
together.  Frozen random-number streams pair the design, errors, tuning data,
prediction sample, and null draws across the $\Delta$ grid.

The CPSS routine accepts disjoint mandatory and optional sets.  Within each
complementary half-sample it projects out the mandatory design, excludes only
documented residualized-zero candidates, and applies LASSO or MCP to the same
eligible optional family.  Its output separates the mandatory set $A_0$, the
selected extension $\widehat E_g$, and
$\widehat A_g=A_0\cup\widehat E_g$, together with selection frequencies,
fit sizes, convergence status, and penalty-path metadata.

\subsection{Software}

The method is implemented in the \texttt{HDMaxShrink} R package; its core
linear-algebra and simulation routines use RcppArmadillo.
The reported DepMap analysis, fixed-core simulation, and honest-selection
audit use frozen package versions 0.6.0, 0.6.1, and 0.6.2, respectively.
DepMap's MCP paths use \texttt{ncvreg} 3.16.0, whereas the selection audit
uses singleton-group MCP in \texttt{grpreg} 3.6.0.  The current package
retains the latter backend.
The local-convexity diagnostic reported for DepMap belongs to the former
backend and is not supplied by the latter.  Reproduction of each study
requires its recorded release and dependencies, not substitution of the
current GitHub version.

\section{Simulation design}
\label{sec:simulation-design}

The numerical study has two deliberately separate components.  The main
experiment supplies the core $A$ and isolates the FM--SM--test--shrinkage
transition.  A 200-replication audit then estimates the additional effect of
learning $A$ by CPSS on an independent sample.  The audit is not pooled with
the main experiment and does not turn its known-core estimand into an oracle
or post-selection claim.

\subsection{Known-core Gaussian experiment}

We generate independent rows $x_i\sim N_p(0,I_p)$ and errors
$\varepsilon_i\sim N(0,1)$.  The primary configuration is

\[
n=100,\qquad p=10000,\qquad p_1=20,\qquad q=9980.
\]

Five sensitivity configurations vary $p$ from 1,000 to 30,000, vary $p_1$
from 10 to 40, or double $(n,p,p_1)$ while preserving $p/n$ and $p_1/n$.
The complete frozen design is reported in the Supplementary Material.  These
finite configurations assess dimensional sensitivity; they do not by
themselves define an asymptotic growth regime.

For $A=\{1,\ldots,p_1\}$, every core coefficient is nonzero.  The ten-entry
pattern

\[
(1.50,1.25,1.00,0.90,0.80,
-1.25,-1.00,-0.90,0.75,-0.75)^{\mathsf T}
\]

is repeated and rescaled so that $\|\beta_A^0\|_2^2=21.52$ for every $p_1$.
One excluded coordinate has coefficient $\Delta$ and the remaining
$q-1$ excluded coefficients are zero.  Hence $H_0:\beta_B^0=0$ holds at
$\Delta=0$ and fails through a sparse one-coordinate departure otherwise.
The primary grid is
$\{0,0.5,1,2,4,8,12,20,40\}$; the five sensitivity configurations use
$\{0,0.5,1,2,4,12,40\}$.

The primary endpoint is Ridge FM; frozen-fold-CV LASSO FM provides a sparse
sensitivity analysis on the same complete design.  Both families share the exact
same SVD submodel refit, maximum statistic, conditional Gaussian draws,
Monte Carlo p-value, $\widehat\kappa$, and PT/S/PS weights.  Ridge and LASSO
penalties are tuned once on an independent null sample and held fixed along
the paired $\Delta$ path.  Thus only the FM endpoint and the direction from
SM to FM differ between the two families.  Ridge-specific risk and plateau
results are not transferred to LASSO; the pointwise interpolation identity
applies to either fitted endpoint.

The primary loss is full-vector squared coefficient error,
$\|\widehat\beta-\beta^0\|_2^2$, on the original coefficient scale.
Conditional-mean prediction loss is evaluated on 1,000 independent Gaussian
covariate vectors without new response noise.  For FM family
$f\in\{R,L\}$, relative performance is

\[
\operatorname{RPE}_{m,f}
=\frac{\overline L_{\FM_f}}{\overline L_m}.
\]

Ratios are formed after averaging paired losses.  Values above one favor
$m$ over its family-specific FM; no ratio is interpreted as uniform
dominance.  The paper profile uses 500 replications in the primary
configuration and 300 in each sensitivity configuration, for 2,000
replications and 44 configuration--departure cells.  Each replication uses
999 conditional Gaussian draws.  Complete MSEs, Monte Carlo errors,
pointwise RPE intervals, support and tuning diagnostics, and all-case figures
are retained in the Supplementary Material and numerical archive.

\subsection{Independent honest-selection audit}
\label{sec:honest-selection-audit-design}

The second experiment contains four scenarios with 50 replications each.
Every replication uses independent selection, analysis, and prediction
samples, with selection size 200, analysis size 100, and $p=1000$.  Four
mandatory coordinates have coefficients
$(1.50,-1.25,1.00,-0.90)^{\mathsf T}$ and two optional active coordinates
have coefficients $(1,-1)^{\mathsf T}$ or
$(0.45,-0.45)^{\mathsf T}$.  Strong and weak signals are crossed with
Gaussian-iid and Toeplitz-$0.5$ designs, so the true support has size six.

CPSS-LASSO and CPSS-MCP use the same 50 complementary half-sample pairs, a
base-selection budget of 10, and stability threshold 0.60.  The mandatory
core is always retained and partialized before optional selection; no
top-ranked fallback is allowed.  The resulting core enters the same
Ridge FM, exact-null SM, conditional Gaussian maximum test, and PT/S/PS
construction as above.  Ridge tuning uses only the selection sample, and
each analysis fit uses 499 conditional Gaussian draws.  This focused audit
measures the honest selection--estimation interface; its 200 replications
remain distinct from the 2,000-replication known-core experiment.

\section{DepMap WRN application}
\label{sec:real-data}

\subsection{Data and frozen honest protocol}

We use the DepMap Public 26Q1 release \citep{depmap26q1} to predict continuous
WRN Chronos gene effect from protein-coding expression and prespecified
biological/design covariates in four MSI-relevant lineages: Bowel,
Esophagus/Stomach, Uterus, and Ovary/Fallopian Tube.  More negative outcomes
indicate stronger dependency.  Five unchanged release files provide model
annotations, CRISPR gene effect, log-transformed protein-coding expression,
global genomic signatures, and the release dictionary.  No missing value is
imputed.  Exclusions and outcome-free removal of nonfinite or zero-variance
expression columns leave $n=211$ observations and $p=19{,}152$ predictors.

A frozen stratified split assigns 82 observations to submodel selection and
Ridge tuning and 129 to analysis.  The split, selected core, and Ridge penalty
are fixed before prediction assessment and are not relearned within its
repeated folds.  MSI score and three lineage indicators form the mandatory
core; Bowel is the reference lineage.  The optional universe contains 19,148
expression predictors before half-sample eligibility checks.

Within each CPSS half-sample the mandatory covariates are partialized before
optional selection.  Primary CPSS-LASSO and sensitivity CPSS-MCP use the same
50 complementary pairs, base-selection budget 10, and threshold 0.60.  An
empty optional extension is retained rather than replaced by a top-ranked
gene.  The Ridge penalty is chosen on the selection sample by a frozen
five-fold split over 121 values from $10^{-4}$ to $10^8$ and then held fixed.
Thus both model construction and tuning are response-separated from the
analysis sample.

\subsection{Prediction and diagnostic sensitivity}

On the 129 analysis observations, all methods are compared on the same held-
out rows using 10 repeats of five-fold cross-validation.  Within each outer
training fold, continuous predictors and the response are centered and
RMS-scaled using training rows only, and the resulting map is applied
unchanged to the test fold.  FM is a single Ridge fit; each SM is a fresh
SVD refit on its fixed mandatory-plus-CPSS core.  PT/S/PS use the corresponding
outer-training maximum statistic.  The primary estimand is repeat-pooled
out-of-fold prediction MSE; RMSE, MAE, and $R^2$ are secondary.  Repeated-fold
dispersion is descriptive because folds are dependent.

The conditional Gaussian maximum test defines the reported shrinkage weights.
Because real-data errors need not be iid homoskedastic Gaussian, we also
compute a studentized Gaussian multiplier maximum as a test-only robustness
diagnostic.  Its decision is not substituted into PT/S/PS after observing the
results.  Agreement would support the Gaussian working calibration; disagreement
is retained as an explicit limitation rather than reconciled by post hoc
weight selection.

\section{Numerical results}
\label{sec:results}

All results below come from completed, manifest-validated frozen artifacts.
The main known-core experiment contains 2,000 replications across six
configurations; the independent honest-selection audit contains 200
replications across four scenarios; and the DepMap analysis uses 10 repeats
of five-fold out-of-fold prediction.  These three experiments have different
estimands and are not pooled.

\subsection{Known-core risk and test transition}

\begin{figure}[t]
\centering
\includegraphics[width=\linewidth]{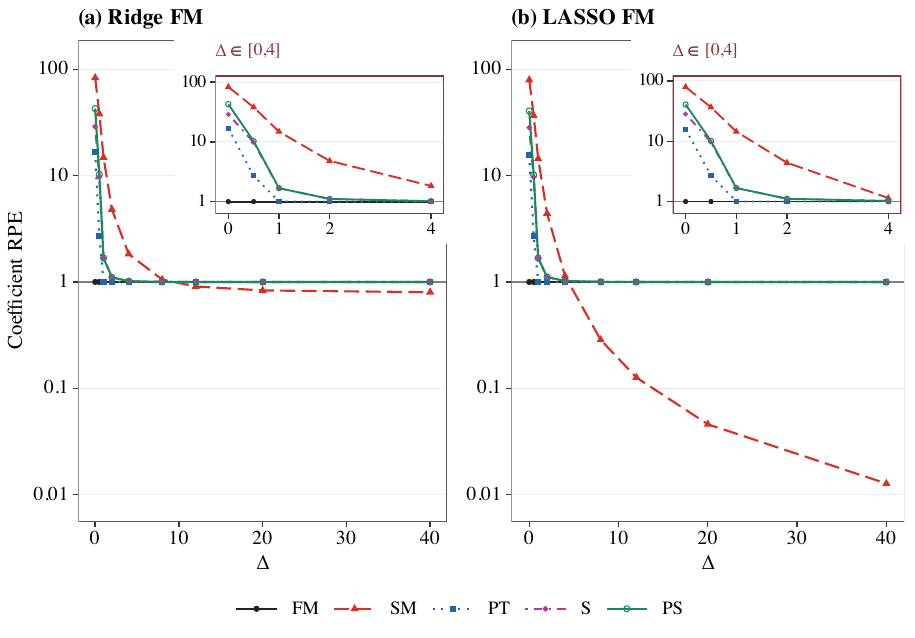}
\caption{Coefficient-risk ratios in the primary known-core configuration:
$n=100$, $p=10000$, $p_1=20$, $q=9980$, and 500 replications.  Each panel
uses its own Ridge/LASSO FM reference.  RPE is
$\mathrm{MSE}(\FM)/\mathrm{MSE}(\text{method})$, so values above one favor
the displayed method.  The vertical scale is logarithmic, the departure axis
is linear through 40, and the insets magnify $0\leq\Delta\leq4$.}
\label{fig:known-core-rpe}
\end{figure}

At $\Delta=0$, SM coefficient MSE is 0.25707, close to its exact
Gaussian-iid value 0.25641.  PS MSE is 0.5010 under Ridge FM and 0.5028 under
LASSO FM, compared with FM MSEs 21.3571 and 20.3695;
the corresponding RPEs are 42.627 and 40.510.  This is a known-core
benchmark, not a statement that a data-selected core must attain the same
gain.  In particular, the null-tuned LASSO FM retains on average only 6.722
of the 20 active core variables.

As the one-coordinate violation increases, SM incurs restriction bias and PS
moves toward FM (Figure~\ref{fig:known-core-rpe}).  At $\Delta=1$, PS RPE
is 1.677 and 1.675 for Ridge and LASSO FMs; at $\Delta=4$ the values are 1.017
and 1.022.  At $\Delta=40$, SM MSE is 2008.204, while Ridge FM and LASSO FM
MSEs are 1608.808 and 25.592.  PS RPE is then 1.000003 and 0.999380,
respectively.  Thus PS approaches either endpoint reference closely but does
not uniformly dominate it.  The positive Ridge/SM tail plateau and the much
smaller LASSO-based SM RPE are distinct endpoint phenomena.

The same pattern persists across the five dimensional sensitivity
configurations and for independent conditional-mean prediction.  At the
null, Ridge-family PS RPE ranges from 24.423 to 62.778 and LASSO-family PS RPE
from 24.996 to 46.366.  Complete all-case coefficient and prediction figures,
analytic SM-risk checks, Monte Carlo intervals, and tuning/support
diagnostics are in the Supplementary Material.  They retain the least
favorable PS coefficient RPE, 0.999076 at $p=30000$ and $\Delta=40$, rather
than rounding it into a dominance claim.

The primary null rejection rate is $24/500=0.048$ at nominal level 0.05.
Across all six configurations, null rates range from 0.0367 to 0.0667 and
every pointwise Wilson interval contains 0.05.  At fixed $(n,p_1)=(100,20)$,
power at $\Delta=0.5$ decreases from 0.5933 to 0.3540 and 0.3067 as $p$
increases from 1,000 to 10,000 and 30,000; the corresponding mean critical
values increase from 4.2879 to 4.8986 and 5.1785.  Primary power is 0.996 at
$\Delta=1$ and equals one from $\Delta=2$ onward.  These are operating
characteristics of the conditionally simulated Gaussian law, not evidence
for a $\chi_q^2$ reference or arbitrary error distributions.  Rejection also
does not set the continuous PS weight exactly to one: its primary mean is
0.999854 at $\Delta=40$.

\FloatBarrier

\subsection{Honest submodel-selection audit}
\label{sec:honest-selection-audit-results}

\begingroup
\setlength{\tabcolsep}{2.3pt}
\renewcommand{\arraystretch}{1.10}
\begin{table}[!htbp]
\footnotesize

\caption{\label{tab:honest-selection-interface}Honest submodel-selection interface audit.}
\centering
\begin{tabular}[t]{llrrrrrrrr}
\toprule
Case & Selector & \shortstack[r]{Sure\\$S_0\subseteq\widehat A$} & \shortstack[r]{Exact\\$\widehat A=S_0$} & $\mathbb E|\widehat A|$ & \shortstack[r]{Selection\\time (s)} & $R_0$ & \shortstack[r]{Rejection\\95\% CI} & \shortstack[r]{SM\\RPE} & \shortstack[r]{PS\\RPE}\\
\midrule
IID-S & LASSO & 1.000 & 0.960 & 6.040 & 2.424 & 50 & 0.060 [0.021, 0.162] & 105.283 & 32.572\\
IID-S & MCP & 1.000 & 0.940 & 6.060 & 42.068 & 50 & 0.060 [0.021, 0.162] & 103.514 & 32.372\\
IID-W & LASSO & 0.880 & 0.840 & 5.920 & 1.986 & 44 & 0.045 [0.013, 0.151] & 87.368 & 34.435\\
IID-W & MCP & 0.900 & 0.860 & 5.940 & 41.620 & 45 & 0.044 [0.012, 0.148] & 86.638 & 34.635\\
T(0.5)-S & LASSO & 1.000 & 0.960 & 6.060 & 2.056 & 50 & 0.080 [0.032, 0.188] & 81.745 & 26.454\\
T(0.5)-S & MCP & 1.000 & 0.960 & 6.040 & 41.875 & 50 & 0.080 [0.032, 0.188] & 80.567 & 26.328\\
T(0.5)-W & LASSO & 0.960 & 0.940 & 5.980 & 1.893 & 48 & 0.083 [0.033, 0.196] & 52.220 & 33.124\\
T(0.5)-W & MCP & 0.900 & 0.880 & 5.920 & 41.147 & 45 & 0.089 [0.035, 0.207] & 52.132 & 33.067\\
\bottomrule
\end{tabular}
\par\vspace{0.35em}
\noindent\begin{minipage}{\linewidth}\footnotesize
\textit{Notes.} Each scenario has 50 replications with an independent selection sample of size 200, an analysis sample of size 100, p = 1000, and 499 conditional Gaussian draws. R0 counts replications for which the selected restriction is truly null; rejection and coefficient RPE are conditional on that stratum. Brackets are pointwise Wilson 95\% Monte Carlo intervals. RPE uses the matched Ridge FM, and a value above one favors the displayed estimator. The weak-signal false-null strata contain only 2--6 replications and are not used for a power claim. The support columns refer to the six-coordinate true active support, not only the four-coordinate mandatory core. S and W denote the strong- and weak-signal regimes; T(0.5) denotes Toeplitz correlation 0.5. This pilot audit is not pooled with the known-core study.
\end{minipage}\par
\end{table}
\endgroup

Table~\ref{tab:honest-selection-interface} reports the separate
200-replication selection audit.  With strong signals, both selectors achieve
sure screening in every replication and exact-support rates of 0.94--0.96.
With weak signals, sure-screening rates are 0.88--0.96 and exact-support
rates are 0.84--0.94; mean false-positive and false-negative counts are
0.02--0.06 and 0--0.12.  The selected submodel is therefore data-adaptive
but remains close to the six-coordinate truth in these focused scenarios.

Conditional on the selected restriction being truly null, rejection rates
range from 0.0444 to 0.0889, and every displayed Wilson interval contains
0.05.  Conditional SM and PS coefficient RPEs are 52.13--105.28 and
26.33--34.63 against the matched Ridge FM.  The weak-signal false-null strata
contain only 2--6 replications per design--selector pair, and the
strong-signal strata contain none; they are therefore not used for a power
claim.  Nor are these conditional gains interpreted as an unconditional
oracle theorem.  CPSS-MCP has similar selection accuracy here but requires
about 41--42 seconds per replication rather than 1.9--2.4 seconds for
CPSS-LASSO.

\FloatBarrier

\subsection{DepMap WRN analysis}

Both CPSS selectors retain the four mandatory variables and select no
optional expression predictor at threshold 0.60.  Hence the LASSO- and
MCP-labeled SM endpoints coincide; their labels do not represent distinct
selected gene sets.  The Ridge penalty, $79{,}432.82$, is selected at grid
index 90 of 121 and is strictly interior to the frozen grid.  The full
selection record is reported in the Supplementary Material.

\begingroup
\setlength{\tabcolsep}{2.3pt}
\renewcommand{\arraystretch}{1.10}
\begin{table}[!htbp]
\small

\caption{\label{tab:depmap-performance}DepMap prediction performance by CPSS selector.}
\centering
\begin{tabular}[t]{lrrrrrr}
\toprule
Estimator & \shortstack[r]{OOF\\PMSE} & \shortstack[r]{Repeat\\SD} & RMSE & MAE & $R^2$ & \shortstack[r]{RPE vs\\Ridge FM}\\
\midrule
\multicolumn{7}{l}{\textbf{Common Ridge FM}}\\*
Ridge FM & 0.4356 & 0.0058 & 0.6600 & 0.4146 & -0.0195 & 1.000000\\
\addlinespace[0.65ex]
\multicolumn{7}{l}{\textbf{Panel A. CPSS-LASSO}}\\*
SM & 0.2170 & 0.0090 & 0.4658 & 0.2541 & 0.4920 & 2.006959\\
PT & 0.3881 & 0.0355 & 0.6224 & 0.3900 & 0.0918 & 1.122518\\
S & 0.3086 & 0.0081 & 0.5555 & 0.3318 & 0.2777 & 1.411491\\
PS & 0.3086 & 0.0081 & 0.5555 & 0.3318 & 0.2777 & 1.411491\\
\addlinespace[0.65ex]
\multicolumn{7}{l}{\textbf{Panel B. CPSS-MCP}}\\*
SM & 0.2170 & 0.0090 & 0.4658 & 0.2541 & 0.4920 & 2.006959\\
PT & 0.3959 & 0.0358 & 0.6286 & 0.3942 & 0.0735 & 1.100306\\
S & 0.3088 & 0.0080 & 0.5557 & 0.3319 & 0.2773 & 1.410623\\
PS & 0.3088 & 0.0080 & 0.5557 & 0.3319 & 0.2773 & 1.410623\\
\bottomrule
\end{tabular}
\par\vspace{0.35em}
\noindent\begin{minipage}{\linewidth}\footnotesize
\textit{Notes.} Repeat SD is descriptive and is not an iid-fold standard error. RPE is Ridge FM PMSE divided by method PMSE; larger is better. All methods use the same held-out observations. The panel labels identify the CPSS selector, not a different full-model or submodel fitting method. Both selectors retain the same four mandatory predictors and no optional extension; their SM estimates coincide. Small shrinkage differences reflect separately seeded Gaussian calibration, not different selected submodels.
\end{minipage}\par
\end{table}
\endgroup

As shown in Table~\ref{tab:depmap-performance}, Ridge FM has repeat-pooled
out-of-fold PMSE 0.4356 and $R^2=-0.0195$.  SM has PMSE 0.2170 and
$R^2=0.4920$ (RPE 2.007).  Primary PS has PMSE 0.3086, $R^2=0.2777$, and
RPE 1.411; primary PT has RPE 1.123.  These are paired held-out prediction
comparisons on the same observations, not coefficient-risk or uniform
dominance claims.

The restriction diagnostics disagree sharply.  Under the Gaussian working
law, the full-analysis statistic is 8.410 against a critical value near 4.9,
with Monte Carlo $p=0.001$ and PS weight about 0.759; the outer fits reject in
86\% and 88\% of the LASSO- and MCP-labeled analyses.  The studentized
multiplier diagnostic gives statistic 3.576 and $p\approx0.46$ and never
rejects in the 50 outer fits of either family.  The robust diagnostic is
test-only and was not substituted into the reported estimator weights.
Complete Gaussian and robust diagnostic tables are retained in the
Supplementary Material.  Their disagreement is evidence that the real-data
calibration is sensitive to the homoskedastic Gaussian assumption, not a
minor numerical discrepancy.

\FloatBarrier

\section{Discussion}
\label{sec:discussion}

The proposed framework makes preliminary-test--Stein-type interpolation operational when
$p\gg n$ and the restriction is data adaptive.  Independent responses select
the submodel; a complete-design regularized estimator and a fresh exact-null
refit define the two endpoints; and one conditional maximum-statistic law
both tests the selected restriction and calibrates the shrinkage weight.  The
link through this common selected-null law, rather than any individual
component, is the central contribution.

The interpretation depends on several distinctions.  The FM is not OLS and
does not invert the singular full Gram matrix.  The SM imposes the selected
coordinate null but need not recover the true support.  Honesty separates the
responses used for selection from those used for estimation and testing; it
does not make the selected restriction correct.  Finally, $q$ counts excluded
coordinates but is neither the number of active variables nor a chi-squared
degree of freedom.  Its classical role is replaced by a design-specific
inverse moment of the simulated maximum-statistic law.

When the selected restriction is approximately correct, the SM can avoid
substantial full-vector variance and Ridge null-space bias, and PS can retain
much of that advantage.  Under a strong departure the test drives PS toward
the corresponding FM\@.  No uniform ordering is claimed between these regimes:
the pointwise interpolation identity can favor an interior combination, but
does not guarantee dominance over both endpoints for every parameter.  The
inverse-moment centering likewise does not import the classical James--Stein
dominance theorem.

The known-core experiments display this endpoint adaptation across six
Gaussian-iid configurations with as many as 30,000 predictors.  At the null
in the primary configuration, PS improves coefficient MSE by factors 42.6
and 40.5 relative to the Ridge and LASSO FMs.  At large
departures, PS risk is close to its family-specific FM risk, with convergence
from either side.  Absolute MSE is therefore needed to compare FM families;
RPE only compares methods within its stated reference family.  Penalties are
tuned on independent null data and then frozen across $\Delta$, so the study
evaluates a reproducible estimator rather than an oracle that retunes for
each departure.  Converged zero LASSO fits and Ridge boundary selections are
retained and reported.

The independent selection-interface audit complements, rather than replaces,
the known-core experiment.  Across its iid and Toeplitz scenarios,
CPSS--LASSO and CPSS--MCP attain sure-screening rates of 0.88--1.00 and
exact-support rates of 0.84--0.96; the selected-null-true pointwise Wilson
intervals contain 0.05 (Section~\ref{sec:honest-selection-audit-results} and
Table~\ref{tab:honest-selection-interface}).  With only 50 replications per
scenario, however, the audit demonstrates the honest interface rather than
uniform support recovery or precise conditional power after rare selection
failures.  MCP shows no accuracy advantage in these settings and is
substantially slower in the recorded implementation.

The DepMap analysis illustrates both the usefulness and the limitation of
the procedure.  On the frozen honest split, the mandatory four-variable SM
predicts better than the Ridge FM fitted to all expression predictors,
and PS lies reproducibly between them.  Both CPSS extensions are empty at the
prespecified threshold, so this is evidence for that particular core, split,
candidate universe, and tuning protocol---not a general claim that small
biological models outperform genomic models.  More importantly, the
conditional-Gaussian diagnostic rejects in most outer fits, whereas the
heteroskedasticity-robust multiplier sensitivity diagnostic does not.  The
application consequently supports predictive interpolation but not a
distribution-free testing claim.

Finite-sample conditional validity requires homoskedastic Gaussian errors.
The multiplier diagnostic is asymptotic and test-only; it does not yet provide
a justified robust inverse-moment weight.  The maximum statistic is also
directed toward sparse departures and can have limited power against many
small violations.  Natural extensions include robust conditional
calibration, a theoretically matched dense-alternative component,
cross-fitting, group or general linear restrictions, non-Gaussian outcomes,
and random-matrix coefficient-risk limits.  Each would require new null-law
and risk analysis rather than a mechanical substitution.

The \texttt{HDMaxShrink} package and accompanying frozen analysis workflow
provide the endpoint solvers, selection interface, null simulation, and code
needed to reproduce the reported tables and figures.

\section*{Data and code availability}
\JASACodeAvailability
The reproducibility materials include frozen configurations, aggregate
outputs, validation records, and scripts for the reported tables and figures.
Raw DepMap files
are obtained from the public DepMap release under its data-access terms and
are not redistributed with the manuscript.

\if1\anon
\section*{Declarations}
\noindent\textit{Funding.} \JASAFundingStatement

\noindent\textit{Competing interests.} \JASACompetingStatement

\noindent\textit{Declaration of generative AI assistance.}
\JASAGenerativeAIStatement
\else
\section*{Disclosure statement}
Funding and competing-interest statements are supplied in the identified
manuscript.
\fi

\clearpage
\appendix
\spacingset{1}
\setcounter{table}{0}
\setcounter{figure}{0}
\renewcommand{\thetable}{S\arabic{table}}
\renewcommand{\thefigure}{S\arabic{figure}}

\section*{Supplementary material}

The following supplementary sections contain the proofs, complete six-configuration
numerical results, selection-interface details, and DepMap robustness diagnostics.

\section{Proofs and technical derivations}
\label{sec:supp-proofs}
\label{sec:appendix-proofs}

\subsection{Proof of Proposition~\ref{prop:sm-risk}}

Because $X_A$ is centered and has full column rank,
$X_A^+X_A=I_{p_1}$ and $X_A^+H_n=X_A^+$.  Substitution of the model into the
SM refit gives
\[
\widehat\beta_A^{\SM}-\beta_A^0
=X_A^+X_B\beta_B^0+X_A^+\varepsilon,
\qquad
\widehat\beta_B^{\SM}-\beta_B^0=-\beta_B^0.
\]
The retained and excluded coordinates occupy disjoint blocks.  The
conditional mean of the noise term is zero and
\[
\E\|X_A^+\varepsilon\|_2^2
=\sigma^2\operatorname{tr}
\{X_A^+(X_A^+)^{\mathsf T}\}
=\sigma^2\|X_A^+\|_F^2,
\]
which proves \eqref{eq:sm-risk}.  If $X_A$ is rank deficient, the same
calculation includes
$(X_A^+X_A-I_{p_1})\beta_A^0$ in the retained-coordinate bias; this is why
the main proposition states the checked rank condition.

\subsection{Proof of Corollary~\ref{cor:sm-iid}}

After removal of the sample mean,
$X_A^{\mathsf T}X_A\sim W_{p_1}(I_{p_1},n-1)$.  Conditional on $X_A$, the
omitted linear predictor $X_B\beta_B^0$ is independent Gaussian noise with
covariance $\|\beta_B^0\|_2^2H_n$.  Therefore
\[
\E\{\|X_A^+X_B\beta_B^0\|_2^2\mid X_A\}
=\|\beta_B^0\|_2^2
\operatorname{tr}\{(X_A^{\mathsf T}X_A)^{-1}\}.
\]
For a Wishart matrix with $n-1$ degrees of freedom,
\[
\E\{(X_A^{\mathsf T}X_A)^{-1}\}
=\frac{I_{p_1}}{n-p_1-2}.
\]
Combining this identity with the error variance in
Proposition~\ref{prop:sm-risk} yields \eqref{eq:sm-risk-iid}.

\subsection{Proof of Proposition~\ref{prop:fm-risk}}

Write $S_s=X_s^{\mathsf T}X_s$.  Since $X_s=X_cD_x^{-1}$,
\[
K_a=S+aD_x^2=D_x(S_s+aI_p)D_x.
\]
An economy SVD of $X_s$ gives the dual identity
\[
(S_s+aI_p)^{-1}X_s^{\mathsf T}
=X_s^{\mathsf T}(X_sX_s^{\mathsf T}+aI_n)^{-1}.
\]
Using $y_s=y_c/s_y$, the implemented backscaled estimator is
\begin{align*}
\widehat\beta^{\FM}
&=s_yD_x^{-1}(S_s+aI_p)^{-1}X_s^{\mathsf T}y_s\\
&=D_x^{-1}(S_s+aI_p)^{-1}D_x^{-1}X_c^{\mathsf T}y_c\\
&=K_a^{-1}X_c^{\mathsf T}y_c.
\end{align*}
Thus response standardization cancels exactly, whereas predictor
standardization remains through $D_x^2$.  Conditional on the estimation
design and the independent selection sample, $D_x$ and $a$ are fixed.
Since $K_a^{-1}S-I_p=-aK_a^{-1}D_x^2$ and
$X_c^{\mathsf T}H_n=X_c^{\mathsf T}$,
\[
\widehat\beta^{\FM}-\beta^0
=-aK_a^{-1}D_x^2\beta^0+K_a^{-1}X_c^{\mathsf T}\varepsilon.
\]
The second term has conditional mean zero and covariance
$\sigma^2K_a^{-1}SK_a^{-1}$.  The squared bias plus the trace of this
covariance proves \eqref{eq:fm-risk}.  This argument does not assume that
$S$ and $D_x$ commute or that $X_c^{\mathsf T}X_c$ is invertible.

\subsection{Proof of Proposition~\ref{prop:large-delta-plateau}}

Substitute $\beta^0(\Delta)=\beta^{(0)}+\Delta v$ into
Proposition~\ref{prop:fm-risk}.  The design, $D_x$, and $a$ remain fixed;
the response scale cancels, so the conditional variance term is independent
of $\Delta$.  Hence
\[
R_{\FM}(X;\Delta)
=\Delta^2a^2\|K_a^{-1}D_x^2v\|_2^2+O(|\Delta|)+O(1).
\]
Likewise, Proposition~\ref{prop:sm-risk} and $v_A=0$ give
\[
R_{\SM}(X,A;\Delta)
=\Delta^2\{\|v_B\|_2^2+
\|X_A^+X_Bv_B\|_2^2\}+O(|\Delta|)+O(1).
\]
Division proves \eqref{eq:large-delta-plateau}.  The denominator is positive
because $v_B\ne0$.  The numerator is positive for fixed $a>0$ because
$K_a$ and $D_x$ are invertible and $v\ne0$.

For the subsequent small-penalty limit, write
\[
aK_a^{-1}D_x^2v
=D_x^{-1}\{a(S_s+aI_p)^{-1}\}D_xv,
\qquad S_s=X_s^{\mathsf T}X_s.
\]
The spectral multiplier $a/(s+a)$ tends to one for a zero eigenvalue of
$S_s$ and to zero for every positive eigenvalue.  Consequently the limit is
$D_x^{-1}P_{\operatorname{null}(X_s)}D_xv$.  Although the middle projector
is Euclidean orthogonal in standardized coordinates, its conjugate is
generally an oblique projection onto $\operatorname{null}(X_c)$ in the
original coefficient metric.  No original-scale Euclidean projection is
substituted for this scale-aware limit.

\subsection{Proof of Theorem~\ref{thm:mc-validity}}

Under $H_0(A)$,
\[
M_Ay=M_AX_A\beta_A^0+M_A\varepsilon=M_A\varepsilon.
\]
Write $\varepsilon=\sigma g_0$, where $g_0\sim N_n(0,I_n)$.  The map
$\mathcal T_A$ is homogeneous of degree zero: multiplying its input by a
positive scalar multiplies both every numerator and the residual-norm scale
by that scalar.  Conditional on $(\mathcal D_S,X)$,
\[
\mathcal T_A(g_0;X),\mathcal T_A(g_1;X),\ldots,
\mathcal T_A(g_{B_{MC}};X)
\]
are therefore iid and hence exchangeable.  The observed statistic has a
uniform rank among the $B_{MC}+1$ values in the absence of ties.  The usual
plus-one p-value is super-uniform with ties, which proves
\eqref{eq:conditional-size}.  The dimension $q$ affects the deterministic map
but not the exchangeability argument.

\subsection{Proof of Corollary~\ref{cor:honest-transfer}}

Condition first on $(\mathcal D_S,X)$.  On every realization for which the
selected null is true, Theorem~\ref{thm:mc-validity} bounds rejection
probability by $\alpha$.  Multiply by the indicator of that event and take
iterated expectations.  Independence of the selection sample and estimation
errors ensures that the conditional Gaussian law used in the theorem is
unchanged by selection.

\subsection{Proof of Proposition~\ref{prop:sphere-representation}}

Because $M_{A_n}=L_nL_n^{\mathsf T}$ and every $z_{nj}$ lies in its range,
\[
r_0=M_{A_n}\varepsilon=\sigma L_nG_n,\qquad
\|r_0\|_2=\sigma R_n,\qquad
\frac{z_{nj}^{\mathsf T}r_0}{\|z_{nj}\|_2}
=\sigma u_{nj}^{\mathsf T}G_n.
\]
It follows that $T_{\rm score}^2=d_nZ_n^2/R_n^2$.  Substitution in the
monotone partial-$t$ transformation gives
\[
\frac{(d_n-1)d_nZ_n^2/R_n^2}
{d_n-d_nZ_n^2/R_n^2}
=\frac{(d_n-1)Z_n^2}{R_n^2-Z_n^2}.
\]
For a finite collection of fixed directions, $Z_n<R_n$ almost surely when
$d_n>1$, so the denominator is positive.  Replacing $\varepsilon/\sigma$ by
an independent standard Gaussian draw proves the calibration-draw identity.

\subsection{Proof of Theorem~\ref{thm:null-width-concentration}}

The map
$g\mapsto\max_j|u_{nj}^{\mathsf T}g|$ is one-Lipschitz.  Gaussian
concentration therefore gives, for every $x>0$,
\[
\Pp_G(|Z_n-a_n|>x\mid X,A_n)\le2\exp(-x^2/2).
\]
Since $a_n\to\infty$, $Z_n/a_n\overset p\longrightarrow1$.  Also
$R_n^2/d_n\overset p\longrightarrow1$ by chi-square concentration.  Using
Proposition~\ref{prop:sphere-representation},
\[
\frac{T_{\max,n}^2}{a_n^2}
=\frac{(1-d_n^{-1})(Z_n^2/a_n^2)}
{R_n^2/d_n-Z_n^2/d_n}
\ \overset p\longrightarrow\ 1,
\]
where $Z_n^2/d_n=(Z_n^2/a_n^2)(a_n^2/d_n)\overset p\longrightarrow0$.

It remains to justify passage to the inverse expectation.  Put
$Y_n=a_n^2/T_{\max,n}^2$.  For $W_n=a_n/Z_n$, the lower-tail condition implies
\[
\sup_n\E_G W_n^8
\le 1+8C\int_1^\infty s^{7-\gamma}\,ds<\infty.
\]
Moreover,
\[
0\le Y_n
=\frac{a_n^2(R_n^2-Z_n^2)}{(d_n-1)Z_n^2}
\le\frac{d_n}{d_n-1}W_n^2\frac{R_n^2}{d_n}.
\]
\begin{samepage}
Cauchy--Schwarz, the displayed inverse eighth-moment bound, and the uniformly
bounded moments of $R_n^2/d_n$ show that $\sup_n\E_GY_n^2<\infty$.  Hence
$Y_n\overset p\longrightarrow1$ is uniformly integrable and $\E_GY_n\to1$.  By definition,
\[
\frac{\kappa_{A_n}(X)}{a_n^2}
=\{\E_GY_n\}^{-1}\longrightarrow1.
\]
\end{samepage}

For the Monte Carlo version, let $Y_{nb}=a_n^2/(T_{\max,n}^{*(b)})^2$.
Conditional on the selected design these variables are iid, and the preceding
second-moment bound is uniform.  Consequently,
\[
\frac1{B_{MC,n}}\sum_{b=1}^{B_{MC,n}}Y_{nb}
-\E_GY_n=o_p(1)
\]
because its conditional variance is $O(B_{MC,n}^{-1})$.  Taking reciprocals
proves $\widehat\kappa_n/a_n^2\overset p\longrightarrow1$.  Slutsky's theorem gives the first
ratio in \eqref{eq:null-calibrated-ratio}.  Finally,
$\epsilon/\widehat\kappa_n\overset p\longrightarrow0$, and
\[
\frac{T_{\max,n}^2\vee\epsilon}{\widehat\kappa_n}
=\max\left\{
\frac{T_{\max,n}^2}{\widehat\kappa_n},
\frac{\epsilon}{\widehat\kappa_n}
\right\}\overset p\longrightarrow1.
\]

\subsection{Proof of Corollary~\ref{cor:iid-max-scale}}

Condition on the selected indices and $X_{A_n}$.  Rotational invariance of
the excluded Gaussian columns gives
\[
L_n^{\mathsf T}X_j\stackrel{\mathrm{iid}}{\sim}N_{d_n}(0,I_{d_n}),
\qquad
u_{nj}\stackrel{\mathrm{iid}}{\sim}
\operatorname{Unif}(\mathbb S^{d_n-1}).
\]
Standard spherical concentration and a union bound yield
\[
\rho_n:=\max_{j\ne k}|u_{nj}^{\mathsf T}u_{nk}|
=O_p\!\left\{\sqrt{\frac{\log(2q_n)}{d_n}}\right\}.
\]
Thus $\rho_n\log(2q_n)\overset p\longrightarrow0$ under
Equation~\eqref{eq:iid-max-rate}.  For fixed directions with $\rho_n<1$,
compare $u_{nj}^{\mathsf T}G_n$ with
$Y_j=\sqrt{1-\rho_n}\,\xi_j+\sqrt{\rho_n}\,\xi_0$, where
$\xi_0,\xi_1,\ldots,\xi_{q_n}$ are independent standard normals.
Their increment variances satisfy
$\E(Y_j-Y_k)^2=2(1-\rho_n)\le
\E_G\{(u_{nj}-u_{nk})^{\mathsf T}G_n\}^2$ for $j\ne k$.
Gaussian comparison and the exponential-moment bound for a maximum give
\begin{equation}
\sqrt{1-\rho_n}\,\E\max_{j\le q_n}\xi_j
\ \le a_n\le\sqrt{2\log(2q_n)}.
\label{eq:iid-width-sandwich}
\end{equation}
The comparison is applied only to the $q_n$ positive-indexed projections;
no independence is claimed for the augmented positive/negative array.
Since $\E\max_{j\le q}\xi_j/\sqrt{2\log q}\to1$,
Equation~\eqref{eq:iid-width-sandwich} proves
$a_n^2/\{2\log(2q_n)\}\overset p\to1$.
In particular $a_n\to\infty$ and $a_n^2/d_n\to0$ on design events with
probability tending to one.  The comparison and concentration tools used
here and below are stated in
\begin{NoHyper}\citet[Lemma~5.33 and Proposition~5.34]{vershynin2012}\end{NoHyper}.

It remains to establish the uniform small-ball bound needed for inverse
moments.  Put $k_n=\lfloor d_n/16\rfloor$; eventually $q_n\ge k_n$.
Represent the first $k_n$ directions as normalized columns of a
$d_n\times k_n$ standard Gaussian matrix $W_n$, independently of $G_n$.
The Gaussian singular-value bound
\begin{NoHyper}\citep[Corollary~5.35]{vershynin2012}\end{NoHyper}, with deviation $\sqrt{d_n}/4$,
and Gaussian concentration for each column norm imply that the event
\[
\mathcal E_n=\{s_{\min}(W_n)\ge\sqrt{d_n}/2,\ 
                 \max_{j\le k_n}\|W_{nj}\|_2\le2\sqrt{d_n}\}
\]
has probability at least $1-2e^{-d_n/32}-k_ne^{-d_n/2}$.
Here $s_{\min}$ denotes the least singular value and $W_{nj}$ the $j$th
column.  For $U_n=(u_{n1},\ldots,u_{nk_n})$, this event implies
$\lambda_{\min}(U_n^{\mathsf T}U_n)\ge c_0:=1/16$.
Conditional on such a design, the Gaussian vector
$U_n^{\mathsf T}G_n$ has a density bounded by
$(2\pi c_0)^{-k_n/2}$.  Integrating that density over the cube
$[-s,s]^{k_n}$ therefore gives the near-zero bound
\begin{equation}
\Pp_G(Z_n\le s\mid X,A_n)\le (C_0s)^{k_n},
\qquad C_0=2/\sqrt{2\pi c_0},\quad s>0.
\label{eq:iid-small-ball-density}
\end{equation}
Equation~\eqref{eq:iid-small-ball-density} bounds a subset event and hence
also the event that all $q_n$ projections are small.
Separately, $G\mapsto\max_j|u_{nj}^{\mathsf T}G|$ is 1-Lipschitz, so
Gaussian concentration yields
\begin{equation}
\Pp_G(Z_n\le ta_n\mid X,A_n)
\le \exp\{-a_n^2(1-t)^2/2\},\qquad 0<t<1.
\label{eq:iid-small-ball-concentration}
\end{equation}
Equation~\eqref{eq:iid-small-ball-concentration} controls deviations below
the conditional mean; the density bound supplies the extra control at zero.

To combine these bounds with constants independent of $n$, fix $\gamma>8$
and set $b=(2C_0)^{-1}$.  For sufficiently large $n$, $k_n>\gamma$ and
$b/a_n<1/2$.  If $0<t<b/a_n$, Equation~\eqref{eq:iid-small-ball-density}
implies
\[
\Pp_G(Z_n\le ta_n\mid X,A_n)
\le 2^{-k_n}(a_n/b)^\gamma t^\gamma\le t^\gamma.
\]
The last inequality holds eventually because $k_n$ is proportional to
$d_n$ and $\log a_n=o(d_n)$.  If $b/a_n\le t\le1/2$,
Equation~\eqref{eq:iid-small-ball-concentration} gives
\[
\Pp_G(Z_n\le ta_n\mid X,A_n)
\le e^{-a_n^2/8}\le(b/a_n)^\gamma\le t^\gamma,
\]
where the middle inequality follows from $a_n\to\infty$.
Finally, for $1/2<t\le1$, the trivial bound $1\le2^\gamma t^\gamma$
applies.  Thus Equation~\eqref{eq:width-small-ball} holds with
$C=2^\gamma$ on the intersection of the preceding high-probability design
events, for all sufficiently large $n$.  Explicitly, for each $0<r<\gamma$,
integration of this bound gives
\[
\E_G\{(a_n/Z_n)^r\mid X,A_n\}
\le 1+\frac{Cr}{\gamma-r}.
\]
Taking $r=8$ supplies the inverse-moment control used in
Theorem~\ref{thm:null-width-concentration}.  Its conditional bounds are
uniform on these design events.  Under the stated null and with
$B_{MC,n}\to\infty$, that theorem and
Equation~\eqref{eq:iid-width-sandwich} therefore yield
Equation~\eqref{eq:iid-tmax-kappa} jointly over design, error, and calibration
draws.  No extreme-value approximation is used by the implemented test.

\subsection{Proof of Proposition~\ref{prop:kappa}}

Conditional on $(X,A)$, the variables $(T_{\max}^{*(b)})^{-2}$ are iid with a
finite second moment.  The strong law of large numbers gives convergence of
their empirical mean to $\E\{(T_0^*)^{-2}\mid X,A\}$.  This limit is finite
and strictly positive under the stated condition.  Continuity of the
reciprocal map proves \eqref{eq:kappa-consistency}.

\subsection{Proof of Proposition~\ref{prop:loss-identity}}

Let $a=\widehat\beta^{\SM}-\beta^0$ and
$b=\widehat\beta^{\FM}-\beta^0$.  Since
$\widehat\beta(w)-\beta^0=(1-w)a+wb$, expansion of the squared norm gives
\[
\|(1-w)a+wb\|_2^2
=(1-w)\|a\|_2^2+w\|b\|_2^2-w(1-w)\|a-b\|_2^2.
\]

\subsection{Proof of Theorem~\ref{thm:separated-alternative}}

In residual coordinates the scaled restricted residual is $G_n+h_n$.  Put
\[
\widetilde Z_n=\max_j|u_{nj}^{\mathsf T}(G_n+h_n)|,
\qquad
\widetilde R_n=\|G_n+h_n\|_2.
\]
The triangle inequality gives
\[
\widetilde Z_n\ge\lambda_n-Z_n.
\]
Gaussian concentration and $a_n\to\infty$ imply $Z_n=O_p(a_n)$.  Condition
\eqref{eq:separated-alternative} in particular implies
$\lambda_n/a_n\to\infty$, and hence
$\widetilde Z_n\ge\lambda_n\{1-o_p(1)\}$.  Also
\[
\widetilde R_n^2=O_p\{d_n+\|h_n\|_2^2\}.
\]
The residual-sphere identity remains valid after replacing $(Z_n,R_n)$ by
$(\widetilde Z_n,\widetilde R_n)$.  Since its denominator is no larger than
$\widetilde R_n^2$,
\[
\frac{T_{\max,n}^2}{a_n^2}
\ge
\frac{(d_n-1)\widetilde Z_n^2}
{a_n^2\widetilde R_n^2}
\overset p\longrightarrow\infty
\]
by \eqref{eq:separated-alternative}.  The calibration draws remain null
Gaussian draws and depend on the design, not on $h_n$, so
$\widehat\kappa_n/a_n^2\overset p\longrightarrow1$.  Division proves
\eqref{eq:alternative-calibrated-ratio}.  If
$\|h_n\|_2^2=O(d_n)$, the left side of
\eqref{eq:separated-alternative} is bounded below, up to a constant, by
$\lambda_n^2/a_n^2$, which proves the simpler sufficient condition.

\subsection{Proof of Theorem~\ref{thm:endpoint-mapping}}

The finite positive calibration and the numerical floor give $V_n>0$.
The function $f(v)=(1-v^{-1})_+$ is continuous at $v=1$ and tends to one
as $v\to\infty$.  Because the definition
\[
V_n=\frac{T_{\max}^2\vee\epsilon}{\widehat\kappa}
\]
retains the guard exactly, the continuous mapping theorem yields
$w_{\PS}=f(V_n)\to0$ in probability under \eqref{eq:null-ratio} and
$w_{\PS}\to1$ under \eqref{eq:alternative-ratio}.  No inactive-guard event
is assumed.

For the stated sufficient conditions involving the raw ratio, observe that
\[
V_n=\max\left\{U_n,\frac{\epsilon}{\widehat\kappa}\right\},
\qquad U_n=\frac{T_{\max}^2}{\widehat\kappa}.
\]
If $U_n\to1$ and $\epsilon/\widehat\kappa\to0$ in probability, continuity
of the maximum gives $V_n\to1$.  If $U_n\to\infty$, the inequality
$V_n\ge U_n$ gives $V_n\to\infty$ without any further guard condition.
The unguarded null ratio alone supplies neither assertion about the floor.

Finally, the endpoint identities are exact:
\[
\|\widehat\beta^{\PS}-\widehat\beta^{\SM}\|_2
=w_{\PS}\|D\|_2,
\qquad
\|\widehat\beta^{\PS}-\widehat\beta^{\FM}\|_2
=(1-w_{\PS})\|D\|_2.
\]
Multiplying $o_p(1)$ by $\|D\|_2=O_p(r_n)$ proves the two $o_p(r_n)$
conclusions.  This does not require independence between the weight and the
endpoint difference.

\subsection{Proof of Corollary~\ref{cor:endpoint-risk-transfer}}

Under the null concentration theorem,
$w_{\PS,n}\overset p\longrightarrow0$.  Since
$e_{\PS,n}=e_{\SM,n}+w_{\PS,n}D_n$ and $0\le w_{\PS,n}\le1$,
\begin{align*}
\big|
\|e_{\PS,n}\|_2^2-\|e_{\SM,n}\|_2^2
\big|
&\le
2w_{\PS,n}\|e_{\SM,n}\|_2\|D_n\|_2
+w_{\PS,n}^2\|D_n\|_2^2\\
&\le
2w_{\PS,n}(\|e_{\SM,n}\|_2+\|D_n\|_2)^2.
\end{align*}
After division by $R_{\SM,n}$, the right side converges to zero in
probability and is uniformly integrable under \eqref{eq:sm-risk-ui}.  Its
expectation therefore tends to zero, proving
$R_{\PS,n}/R_{\SM,n}\to1$.

Under the separated alternative, let
$\delta_n=1-w_{\PS,n}\overset p\longrightarrow0$.  Now
$e_{\PS,n}=e_{\FM,n}-\delta_nD_n$, and the identical inequality with
$(e_{\SM,n},w_{\PS,n},R_{\SM,n})$ replaced by
$(e_{\FM,n},\delta_n,R_{\FM,n})$ proves the second claim from
\eqref{eq:fm-risk-ui}.  Neither argument provides a one-sided risk
inequality.

\subsection{Proof of Proposition~\ref{prop:selection-remainder}}

Apply Cauchy--Schwarz to the nonnegative squared loss and the indicator of
$\mathcal E_n^c$:
\[
\E\{L_2\ind(\mathcal E_n^c)\}
\le(\E L_2^2)^{1/2}
(\E\ind(\mathcal E_n^c)^2)^{1/2}
=(\E L_2^2)^{1/2}\Pp(\mathcal E_n^c)^{1/2}.
\]

\section{Additional theoretical targets}
\label{sec:appendix-targets}

Corollary~\ref{cor:iid-max-scale} proves a $2\log q$ expansion only for the
declared Gaussian-iid regime.  Extending it to structured dependent designs
would require a model-specific effective multiplicity and corresponding
lower-tail verification; nominal $q$ is not universal.  A separate remaining
target is a deterministic equivalent for Ridge coefficient risk, which
requires a declared signal model in addition to spectral convergence of the
design covariance.  Neither extension is needed for finite-sample test
validity or for the exact endpoint decompositions.

\clearpage
\section{Complete simulation and computational diagnostics}
\label{sec:appendix-simulation}

All tables and figures in this supplement are generated from completed,
validated artifacts.  Pilot, legacy single-configuration, and paper-profile
replications are not pooled.  The complete archive retains 1,056 risk records
(528 per loss criterion); the five-estimator presentation uses 440 per
criterion, with the same SM under both FM references.  The figures below
retain every case and departure for FM, SM, PT, S, and PS, while full MSEs,
Monte Carlo errors, and intervals remain in the numerical CSV records.

\subsection{Frozen known-core configurations}

The main experiment fixes the core by design and therefore does not estimate
selection error.  Table~\ref{tab:ultra-hd-scenario-design} gives all six
configurations, replication counts, departure grids, and Monte Carlo sizes.
It is distinct from the four-scenario, 200-replication honest-selection audit
reported in the main article.

\begingroup
\setlength{\tabcolsep}{2.3pt}
\renewcommand{\arraystretch}{1.10}
\begin{table}[!htbp]
\small

\caption{\label{tab:ultra-hd-scenario-design}Ultra-high-dimensional simulation design.}
\centering
\begin{tabular}[t]{lrrrrrrr}
\toprule
Case & $n$ & $p$ & $p_1$ & $q$ & $R$ & $B$ & $|\mathcal D|$\\
\midrule
C1 & 100 & 10000 & 20 & 9980 & 500 & 999 & 9\\
C2 & 100 & 1000 & 20 & 980 & 300 & 999 & 7\\
C3 & 100 & 30000 & 20 & 29980 & 300 & 999 & 7\\
C4 & 100 & 10000 & 10 & 9990 & 300 & 999 & 7\\
C5 & 100 & 10000 & 40 & 9960 & 300 & 999 & 7\\
C6 & 200 & 20000 & 40 & 19960 & 300 & 999 & 7\\
\bottomrule
\end{tabular}
\par\vspace{0.35em}
\noindent\begin{minipage}{\linewidth}\footnotesize
\textit{Notes.} C1 uses $\Delta$ = 0, 0.5, 1, 2, 4, 8, 12, 20, 40; C2--C6 omit 8 and 20. Every core coefficient is active; the core squared norm is 21.52.
\end{minipage}\par
\end{table}
\endgroup

\subsection{Complete coefficient-risk figures}

\begin{figure}[htbp]
\centering
\includegraphics[width=\linewidth]{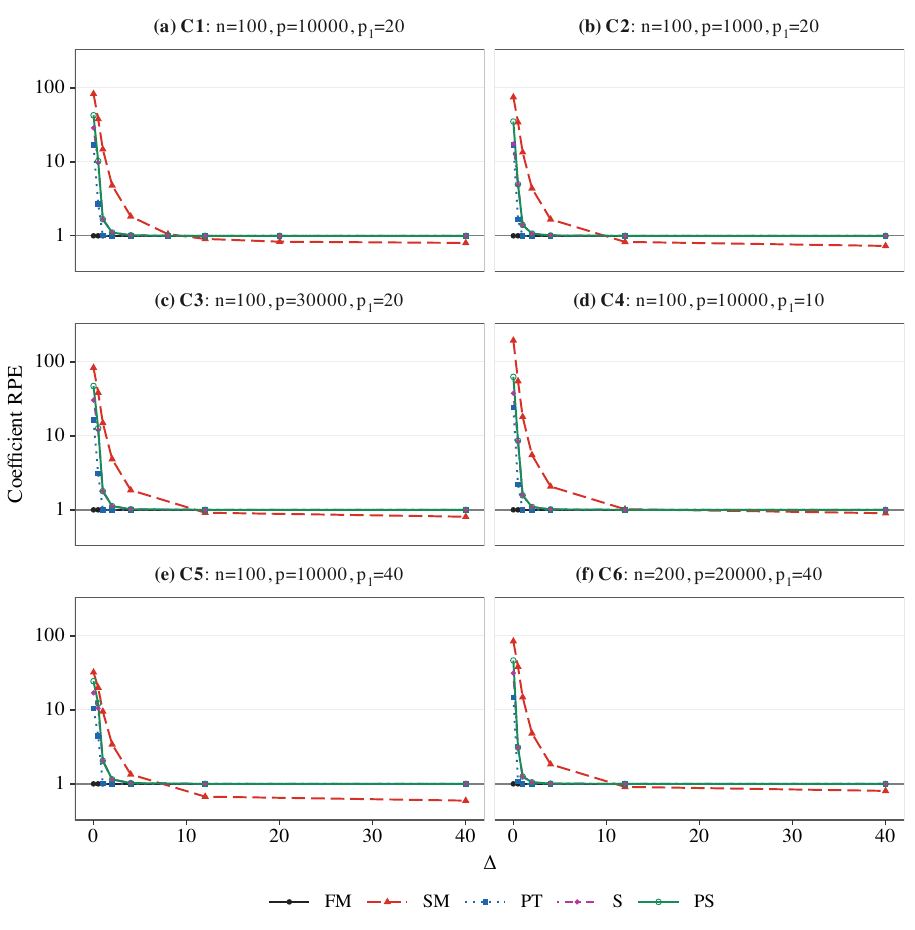}
\caption{Full-vector coefficient RPE for all six cases with a Ridge FM.
RPE is $\mathrm{MSE}(\FM)/\mathrm{MSE}(\text{method})$, with no division
of coefficient loss by $p$.  C1 has 500 replications and C2--C6 have 300 each.
All five estimators and the full $0\leq\Delta\leq40$ path are retained;
coincident curves overlap.  No RPE confidence bands are drawn.}
\label{fig:appendix-all-cases-ridge}
\end{figure}

\begin{figure}[p]
\centering
\includegraphics[width=\linewidth]{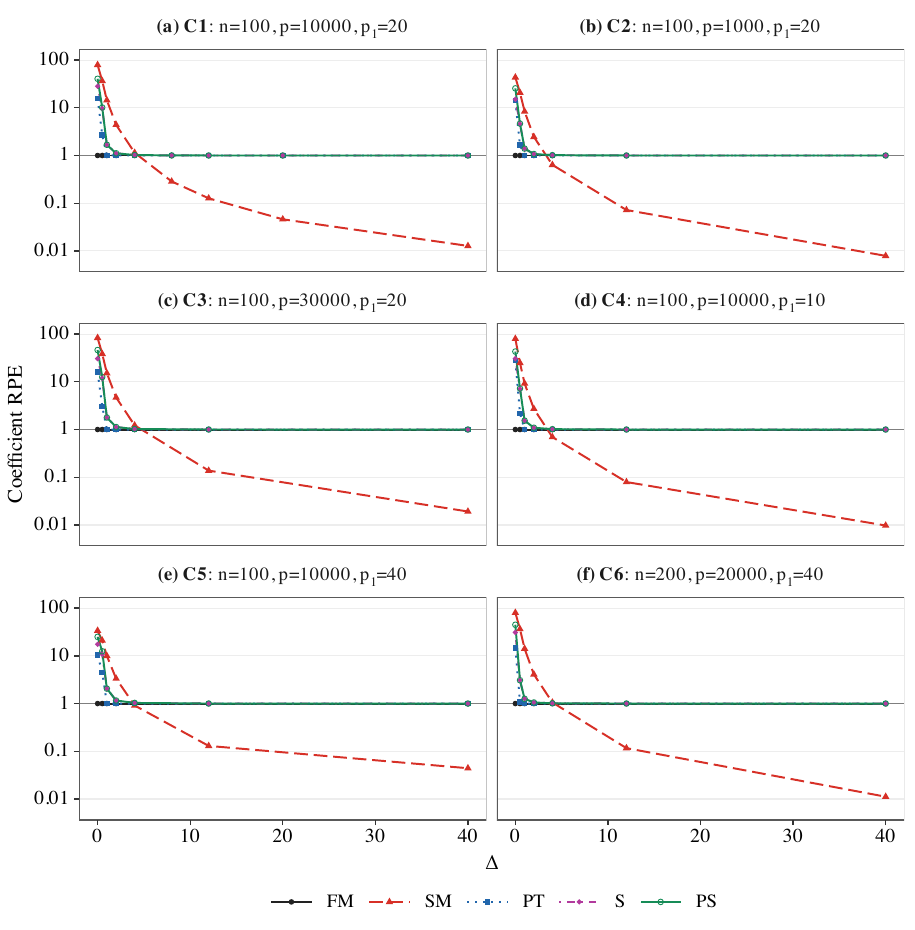}
\caption{Full-vector coefficient RPE for the same six cases with a
LASSO FM.  Each ratio uses its own case-specific LASSO reference, while
SM, the test, and shrinkage calibration are shared with the Ridge family.
All case--departure coordinates for the five methods are retained, including
small PS deficits below one and the zero-support LASSO fits discussed in
the main article.
All $p_1$ core coefficients are active; their total squared norm is 21.52.}
\label{fig:appendix-all-cases-lasso}
\end{figure}

Figures~\ref{fig:appendix-all-cases-ridge} and~\ref{fig:appendix-all-cases-lasso}
compare coefficient RPE over the same six designs using Ridge FM and LASSO FM
references, respectively, isolating how the FM endpoint changes the risk ratios.

\clearpage
\clearpage

\subsection{Prediction risk and Monte Carlo precision}

Coefficient loss is $\|\widehat\beta-\beta\|_2^2$ over all $p$ coordinates.
Figure~\ref{fig:appendix-prediction-rpe} instead reports prediction of the
conditional mean on an independent Gaussian design, without additional
response noise.  The Ridge FM and LASSO FM parts both show large near-null
gains and approach to their own FM references as the restriction becomes
strongly false.  These losses are distinct from observed-response PMSE in
the DepMap analysis.  With paired replication losses $L_{r,F}$ and
$L_{r,m}$, the reported ratio and its Monte Carlo standard error are
\[
\widehat{\mathrm{RPE}}_m=\frac{\overline L_F}{\overline L_m},
\qquad
\widehat{\mathrm{MCSE}}(\widehat{\mathrm{RPE}}_m)
=\frac{\operatorname{sd}\{L_{r,F}
-\widehat{\mathrm{RPE}}_m L_{r,m}\}}
{\sqrt{R}\,\overline L_m}.
\]
The archived 95\% Monte Carlo intervals use
$\widehat{\mathrm{RPE}}\pm t_{R-1,.975}\widehat{\mathrm{MCSE}}$.
They are pointwise, approximate intervals for Monte Carlo uncertainty, not
simultaneous confidence bands or intervals for a real-data coefficient.
No truncation at one is applied.  Full-precision numerical values and paired
standard errors are retained in the accompanying CSV files.

\clearpage
\begin{figure}[p]
\centering
\includegraphics[width=\linewidth]{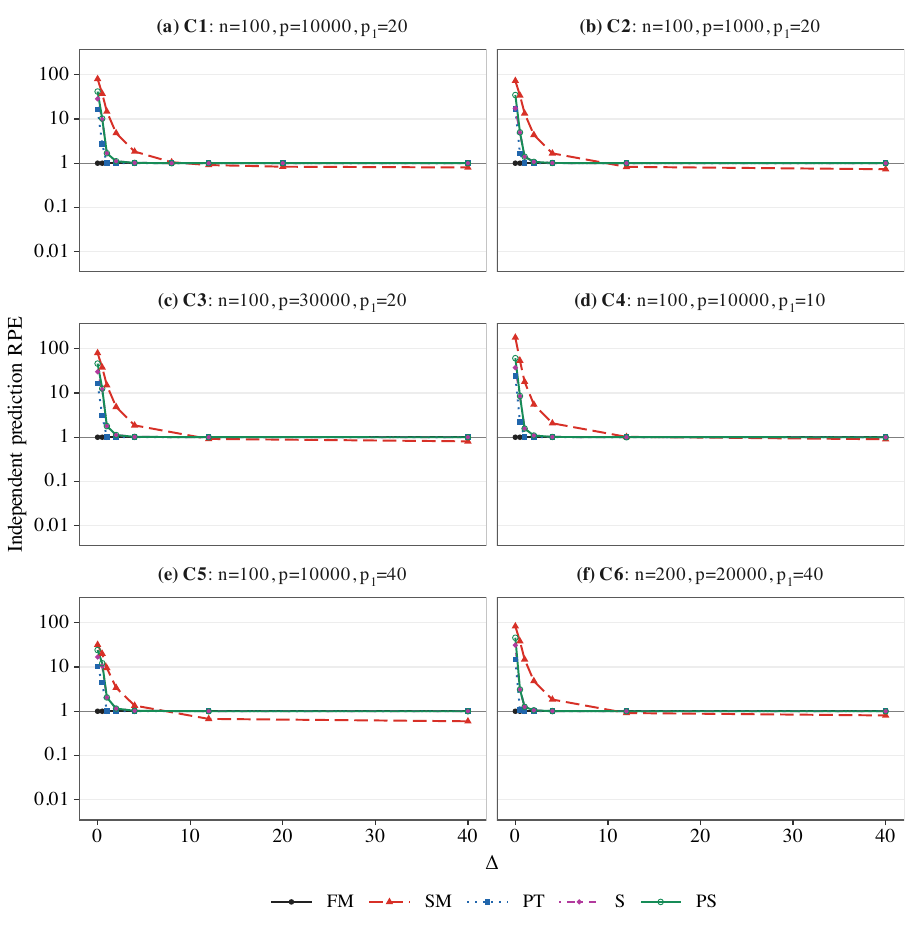}
\caption{Independent conditional-mean prediction RPE in all six cases.
Ridge FM reference.  Each panel retains FM, SM, PT, S, and PS on
the full $0\leq\Delta\leq40$ grid, with 500 replications in C1 and 300 in
C2--C6.  Prediction uses 1,000 independent Gaussian covariate vectors per
replication, including the reconstructed intercept but no new response
noise.  The vertical axis is logarithmic; no confidence bands are drawn.}
\label{fig:appendix-prediction-rpe}
\end{figure}
\clearpage
\begin{figure}[p]
\ContinuedFloat
\centering
\includegraphics[width=\linewidth]{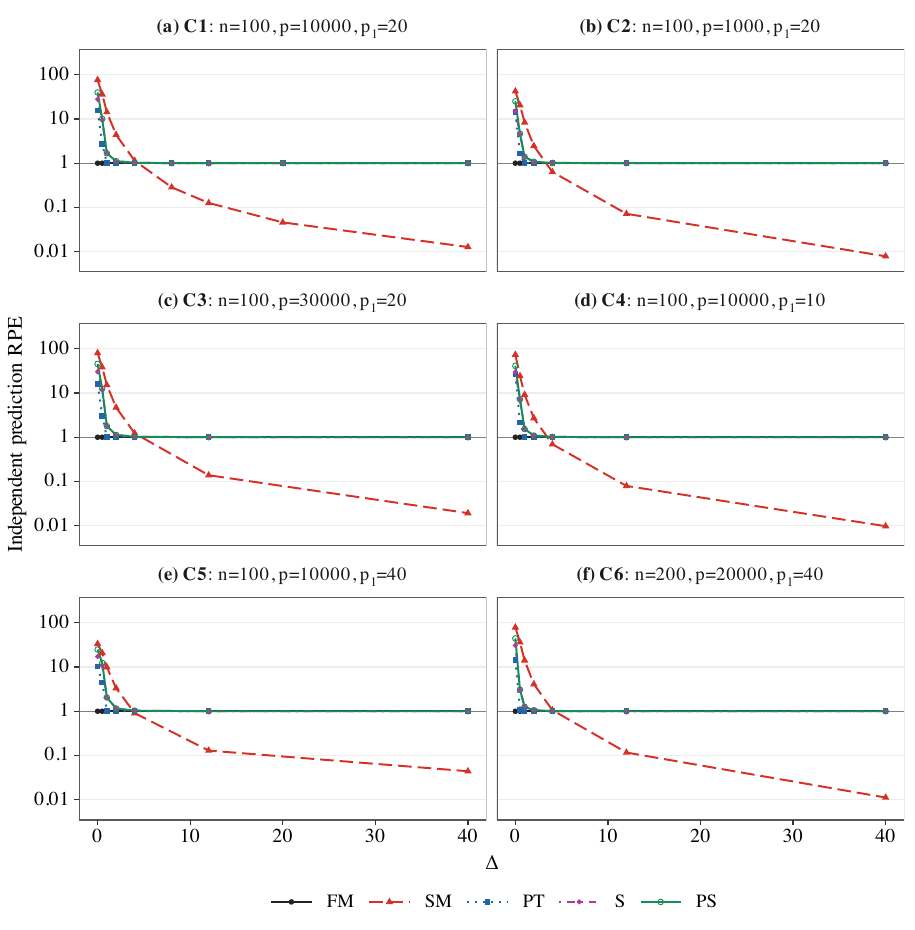}
\caption[]{Independent conditional-mean prediction RPE (continued).
LASSO FM reference.  The designs, SM, test, and scalar
shrinkage weights are shared with the preceding Ridge FM part; only the FM endpoint and its
reference risk change.  Ratios slightly below one are preserved, including
the primary PS ratio 0.999375 at $\Delta=40$.  Coincident curves overlap.}
\end{figure}
\clearpage

\subsection{Test, endpoint, and timing diagnostics}

Table~\ref{tab:test-diagnostics} retains all case--departure combinations.
Wilson intervals
quantify binomial Monte Carlo uncertainty in rejection rates; at $\Delta=0$
all six intervals contain .05, while the C5 estimate itself is .0667.
All configurations use $B_{MC}=999$, so the smallest rank p-value is .001.
The cases share seed streams and should not be pooled as 2,000 independent
null experiments.

Table~\ref{tab:fm-tuning} reports the independently chosen and subsequently
frozen penalties, boundary-selection rates, and validation losses.  The
boundary rates describe the limitations of this tuning protocol; they are
not evidence that the optimization failed.

\begingroup
\fontsize{9}{11}\selectfont
\setlength{\tabcolsep}{2.3pt}
\begingroup
\setlength{\tabcolsep}{2.3pt}
\renewcommand{\arraystretch}{1.10}
\setlength{\LTleft}{0pt plus 1fil}
\setlength{\LTright}{0pt plus 1fil}
\footnotesize

\begin{longtable}{@{\extracolsep{\fill}}lrrrrrrrr@{}}
\caption{\label{tab:test-diagnostics}Maximum-test operating characteristics.}\\
\toprule
Case & $\Delta$ & $R$ & \shortstack[r]{Rejection\\rate} & \shortstack[r]{Rejection\\MC95 interval} & $\overline T_{\max}$ & $\overline c_{1-\alpha}$ & $\overline\kappa$ & $\overline w_{\mathrm{PS}}$\\
\midrule
\endfirsthead
\multicolumn{9}{@{}l}{\textit{\tablename~\thetable. Maximum-test operating characteristics. (continued)}}\\
\toprule
Case & $\Delta$ & $R$ & \shortstack[r]{Rejection\\rate} & \shortstack[r]{Rejection\\MC95 interval} & $\overline T_{\max}$ & $\overline c_{1-\alpha}$ & $\overline\kappa$ & $\overline w_{\mathrm{PS}}$\\
\midrule
\endhead
\midrule
\multicolumn{9}{r}{Continued on next page}\\
\endfoot
\endlastfoot
\multicolumn{9}{@{}l}{\textbf{Case C1}}\\*
C1 & 0.0 & 500 & 0.0480 & {}[0.0325, 0.0704] & 4.269 & 4.899 & 17.764 & 0.062632\\
C1 & 0.5 & 500 & 0.3540 & {}[0.3133, 0.3969] & 4.795 & 4.899 & 17.764 & 0.199909\\
C1 & 1.0 & 500 & 0.9960 & {}[0.9855, 0.9989] & 8.837 & 4.899 & 17.764 & 0.753889\\
C1 & 2.0 & 500 & 1.0000 & {}[0.9924, 1.0000] & 17.715 & 4.899 & 17.764 & 0.940604\\
C1 & 4.0 & 500 & 1.0000 & {}[0.9924, 1.0000] & 35.472 & 4.899 & 17.764 & 0.985297\\
C1 & 8.0 & 500 & 1.0000 & {}[0.9924, 1.0000] & 70.985 & 4.899 & 17.764 & 0.996335\\
C1 & 12.0 & 500 & 1.0000 & {}[0.9924, 1.0000] & 106.499 & 4.899 & 17.764 & 0.998372\\
C1 & 20.0 & 500 & 1.0000 & {}[0.9924, 1.0000] & 177.526 & 4.899 & 17.764 & 0.999414\\
C1 & 40.0 & 500 & 1.0000 & {}[0.9924, 1.0000] & 355.095 & 4.899 & 17.764 & 0.999854\\
\addlinespace[0.4ex]
\multicolumn{9}{@{}l}{\textbf{Case C2}}\\*
C2 & 0.0 & 300 & 0.0467 & {}[0.0280, 0.0768] & 3.537 & 4.288 & 12.422 & 0.070710\\
C2 & 0.5 & 300 & 0.5933 & {}[0.5369, 0.6474] & 4.659 & 4.288 & 12.422 & 0.367466\\
C2 & 1.0 & 300 & 1.0000 & {}[0.9874, 1.0000] & 9.002 & 4.288 & 12.422 & 0.834417\\
C2 & 2.0 & 300 & 1.0000 & {}[0.9874, 1.0000] & 17.960 & 4.288 & 12.422 & 0.959684\\
C2 & 4.0 & 300 & 1.0000 & {}[0.9874, 1.0000] & 35.878 & 4.288 & 12.422 & 0.989972\\
C2 & 12.0 & 300 & 1.0000 & {}[0.9874, 1.0000] & 107.548 & 4.288 & 12.422 & 0.998886\\
C2 & 40.0 & 300 & 1.0000 & {}[0.9874, 1.0000] & 358.393 & 4.288 & 12.422 & 0.999900\\
\addlinespace[0.4ex]
\multicolumn{9}{@{}l}{\textbf{Case C3}}\\*
C3 & 0.0 & 300 & 0.0500 & {}[0.0305, 0.0808] & 4.541 & 5.179 & 20.425 & 0.053846\\
C3 & 0.5 & 300 & 0.3067 & {}[0.2572, 0.3610] & 5.000 & 5.179 & 20.425 & 0.166562\\
C3 & 1.0 & 300 & 1.0000 & {}[0.9874, 1.0000] & 9.006 & 5.179 & 20.425 & 0.726799\\
C3 & 2.0 & 300 & 1.0000 & {}[0.9874, 1.0000] & 18.011 & 5.179 & 20.425 & 0.933666\\
C3 & 4.0 & 300 & 1.0000 & {}[0.9874, 1.0000] & 36.020 & 5.179 & 20.425 & 0.983539\\
C3 & 12.0 & 300 & 1.0000 & {}[0.9874, 1.0000] & 108.056 & 5.179 & 20.425 & 0.998175\\
C3 & 40.0 & 300 & 1.0000 & {}[0.9874, 1.0000] & 360.182 & 5.179 & 20.425 & 0.999836\\
\addlinespace[0.4ex]
\multicolumn{9}{@{}l}{\textbf{Case C4}}\\*
C4 & 0.0 & 300 & 0.0367 & {}[0.0206, 0.0645] & 4.238 & 4.858 & 17.542 & 0.062343\\
C4 & 0.5 & 300 & 0.4433 & {}[0.3882, 0.4999] & 4.961 & 4.858 & 17.542 & 0.246266\\
C4 & 1.0 & 300 & 1.0000 & {}[0.9874, 1.0000] & 9.406 & 4.858 & 17.542 & 0.783469\\
C4 & 2.0 & 300 & 1.0000 & {}[0.9874, 1.0000] & 18.902 & 4.858 & 17.542 & 0.948180\\
C4 & 4.0 & 300 & 1.0000 & {}[0.9874, 1.0000] & 37.894 & 4.858 & 17.542 & 0.987226\\
C4 & 12.0 & 300 & 1.0000 & {}[0.9874, 1.0000] & 113.863 & 4.858 & 17.542 & 0.998590\\
C4 & 40.0 & 300 & 1.0000 & {}[0.9874, 1.0000] & 379.756 & 4.858 & 17.542 & 0.999873\\
\addlinespace[0.4ex]
\multicolumn{9}{@{}l}{\textbf{Case C5}}\\*
C5 & 0.0 & 300 & 0.0667 & {}[0.0436, 0.1007] & 4.319 & 5.024 & 18.468 & 0.057729\\
C5 & 0.5 & 300 & 0.1833 & {}[0.1436, 0.2310] & 4.573 & 5.024 & 18.468 & 0.125470\\
C5 & 1.0 & 300 & 0.9933 & {}[0.9760, 0.9982] & 7.742 & 5.024 & 18.468 & 0.662108\\
C5 & 2.0 & 300 & 1.0000 & {}[0.9874, 1.0000] & 15.491 & 5.024 & 18.468 & 0.918706\\
C5 & 4.0 & 300 & 1.0000 & {}[0.9874, 1.0000] & 30.991 & 5.024 & 18.468 & 0.979850\\
C5 & 12.0 & 300 & 1.0000 & {}[0.9874, 1.0000] & 92.990 & 5.024 & 18.468 & 0.997765\\
C5 & 40.0 & 300 & 1.0000 & {}[0.9874, 1.0000] & 309.989 & 5.024 & 18.468 & 0.999799\\
\addlinespace[0.4ex]
\multicolumn{9}{@{}l}{\textbf{Case C6}}\\*
C6 & 0.0 & 300 & 0.0567 & {}[0.0357, 0.0889] & 4.324 & 4.885 & 18.283 & 0.061633\\
C6 & 0.5 & 300 & 0.9300 & {}[0.8954, 0.9538] & 6.532 & 4.885 & 18.283 & 0.530507\\
C6 & 1.0 & 300 & 1.0000 & {}[0.9874, 1.0000] & 12.929 & 4.885 & 18.283 & 0.886516\\
C6 & 2.0 & 300 & 1.0000 & {}[0.9874, 1.0000] & 25.731 & 4.885 & 18.283 & 0.971738\\
C6 & 4.0 & 300 & 1.0000 & {}[0.9874, 1.0000] & 51.335 & 4.885 & 18.283 & 0.992923\\
C6 & 12.0 & 300 & 1.0000 & {}[0.9874, 1.0000] & 153.754 & 4.885 & 18.283 & 0.999212\\*
C6 & 40.0 & 300 & 1.0000 & {}[0.9874, 1.0000] & 512.219 & 4.885 & 18.283 & 0.999929\\*
\multicolumn{9}{@{}l@{}}{\makebox[0pt][l]{\begin{minipage}[t]{\linewidth}\vspace{0.2em}\hrule height\heavyrulewidth\vspace{0.35em}\footnotesize\textit{Notes.} Intervals are pointwise Wilson Monte Carlo intervals for rejection rates; the statistic, critical value, kappa and weights are replication means.\par\end{minipage}}}\\
\end{longtable}\par
\endgroup

\begingroup
\setlength{\tabcolsep}{2.3pt}
\renewcommand{\arraystretch}{1.10}
\setlength{\LTleft}{0pt plus 1fil}
\setlength{\LTright}{0pt plus 1fil}
\footnotesize

\begin{longtable}{@{\extracolsep{\fill}}llrrrrrrr@{}}
\caption{\label{tab:fm-tuning}Frozen full-model tuning diagnostics.}\\
\toprule
Case & FM & $R$ & \shortstack[r]{Mean\\$\lambda$} & \shortstack[r]{Median\\$\lambda$} & Lower & Upper & \shortstack[r]{CV\\MSE} & Warnings\\
\midrule
\endfirsthead
\multicolumn{9}{@{}l}{\textit{\tablename~\thetable. Frozen full-model tuning diagnostics. (continued)}}\\
\toprule
Case & FM & $R$ & \shortstack[r]{Mean\\$\lambda$} & \shortstack[r]{Median\\$\lambda$} & Lower & Upper & \shortstack[r]{CV\\MSE} & Warnings\\
\midrule
\endhead
\midrule
\multicolumn{9}{r}{Continued on next page}\\
\endfoot
\endlastfoot
\multicolumn{9}{@{}l}{\textbf{Case C1}}\\*
C1 & R & 500 & 1.259e+03 & 5.661e+00 & 0.1560 & 0.1200 & 22.3184 & 0\\
C1 & L & 500 & 3.803e-02 & 3.162e-02 & 0.0520 & 0.0000 & 12.8706 & 0\\
\addlinespace[0.4ex]
\multicolumn{9}{@{}l}{\textbf{Case C2}}\\*
C2 & R & 300 & 1.036e+00 & 5.012e-01 & 0.3533 & 0.0000 & 19.4021 & 0\\
C2 & L & 300 & 1.608e-02 & 1.468e-02 & 0.0033 & 0.0000 & 2.9589 & 0\\
\addlinespace[0.4ex]
\multicolumn{9}{@{}l}{\textbf{Case C3}}\\*
C3 & R & 300 & 2.605e+03 & 1.995e-07 & 0.0000 & 0.2467 & 22.5783 & 0\\
C3 & L & 300 & 8.406e-02 & 8.254e-02 & 0.0167 & 0.0000 & 17.0553 & 0\\
\addlinespace[0.4ex]
\multicolumn{9}{@{}l}{\textbf{Case C4}}\\*
C4 & R & 300 & 1.334e+03 & 1.259e+01 & 0.1533 & 0.1233 & 22.3048 & 0\\
C4 & L & 300 & 2.770e-02 & 3.162e-02 & 0.0033 & 0.0000 & 2.4520 & 0\\
\addlinespace[0.4ex]
\multicolumn{9}{@{}l}{\textbf{Case C5}}\\*
C5 & R & 300 & 1.113e+03 & 1.000e+01 & 0.1300 & 0.1067 & 22.3104 & 0\\
C5 & L & 300 & 5.273e-01 & 1.468e-01 & 0.0033 & 0.0400 & 20.4479 & 0\\
\addlinespace[0.4ex]
\multicolumn{9}{@{}l}{\textbf{Case C6}}\\*
C6 & R & 300 & 5.379e+02 & 1.627e+00 & 0.0800 & 0.0500 & 22.2201 & 0\\*
C6 & L & 300 & 2.569e-02 & 2.610e-02 & 0.0100 & 0.0000 & 13.0574 & 0\\*
\multicolumn{9}{@{}l@{}}{\makebox[0pt][l]{\begin{minipage}[t]{\linewidth}\vspace{0.2em}\hrule height\heavyrulewidth\vspace{0.35em}\footnotesize\textit{Notes.} Penalties are selected on an independent null sample and frozen across $\Delta$. R/L denote Ridge/LASSO FM; Lower and Upper are grid-boundary selection proportions. Boundary and all-zero solutions are retained.\par\end{minipage}}}\\
\end{longtable}\par
\endgroup

\endgroup

Table~\ref{tab:ultra-hd-exact-sm-risk} compares empirical coefficient risk
with the exact SM benchmark
$\Delta^2+(1+\Delta^2)p_1/(n-p_1-2)$.  The largest absolute standardized
empirical discrepancy is 2.41 Monte Carlo standard errors (C5, $\Delta=0$);
the adjacent $\Delta=.5$ cell is 2.24 standard errors below the benchmark.
These are correlated comparisons across the same replications, not evidence
that every cell agrees within a prespecified two-standard-error tolerance.

\begingroup
\fontsize{9}{11}\selectfont
\setlength{\tabcolsep}{2.3pt}
\begingroup
\setlength{\tabcolsep}{2.3pt}
\renewcommand{\arraystretch}{1.10}
\setlength{\LTleft}{0pt plus 1fil}
\setlength{\LTright}{0pt plus 1fil}
\footnotesize

\begin{longtable}{@{\extracolsep{\fill}}lrrrrrrr@{}}
\caption{\label{tab:ultra-hd-exact-sm-risk}Exact submodel-risk check.}\\
\toprule
Case & $\Delta$ & $R$ & Empirical & MCSE & Exact & Ratio & \shortstack[r]{Difference\\in MCSE units}\\
\midrule
\endfirsthead
\multicolumn{8}{@{}l}{\textit{\tablename~\thetable. Exact submodel-risk check. (continued)}}\\
\toprule
Case & $\Delta$ & $R$ & Empirical & MCSE & Exact & Ratio & \shortstack[r]{Difference\\in MCSE units}\\
\midrule
\endhead
\midrule
\multicolumn{8}{r}{Continued on next page}\\
\endfoot
\endlastfoot
\multicolumn{8}{@{}l}{\textbf{Case C1}}\\*
C1 & 0.0 & 500 & 0.2571 & 0.0041 & 0.2564 & 1.002572 & 0.159\\
C1 & 0.5 & 500 & 0.5699 & 0.0049 & 0.5705 & 0.998903 & -0.127\\
C1 & 1.0 & 500 & 1.5102 & 0.0075 & 1.5128 & 0.998275 & -0.347\\
C1 & 2.0 & 500 & 5.2734 & 0.0185 & 5.2821 & 0.998358 & -0.469\\
C1 & 4.0 & 500 & 20.3298 & 0.0640 & 20.3590 & 0.998567 & -0.456\\
C1 & 8.0 & 500 & 80.5630 & 0.2498 & 80.6667 & 0.998714 & -0.415\\
C1 & 12.0 & 500 & 180.9565 & 0.5615 & 181.1795 & 0.998770 & -0.397\\
C1 & 20.0 & 500 & 502.2250 & 1.5629 & 502.8205 & 0.998816 & -0.381\\
C1 & 40.0 & 500 & 2008.2038 & 6.2706 & 2010.5128 & 0.998852 & -0.368\\
\addlinespace[0.4ex]
\multicolumn{8}{@{}l}{\textbf{Case C2}}\\*
C2 & 0.0 & 300 & 0.2613 & 0.0055 & 0.2564 & 1.019118 & 0.897\\
C2 & 0.5 & 300 & 0.5777 & 0.0069 & 0.5705 & 1.012545 & 1.031\\
C2 & 1.0 & 300 & 1.5215 & 0.0114 & 1.5128 & 1.005750 & 0.762\\
C2 & 2.0 & 300 & 5.2917 & 0.0285 & 5.2821 & 1.001825 & 0.338\\
C2 & 4.0 & 300 & 20.3619 & 0.0937 & 20.3590 & 1.000145 & 0.032\\
C2 & 12.0 & 300 & 181.0416 & 0.7696 & 181.1795 & 0.999239 & -0.179\\
C2 & 40.0 & 300 & 2008.4432 & 8.3700 & 2010.5128 & 0.998971 & -0.247\\
\addlinespace[0.4ex]
\multicolumn{8}{@{}l}{\textbf{Case C3}}\\*
C3 & 0.0 & 300 & 0.2575 & 0.0055 & 0.2564 & 1.004248 & 0.200\\
C3 & 0.5 & 300 & 0.5667 & 0.0067 & 0.5705 & 0.993362 & -0.562\\
C3 & 1.0 & 300 & 1.5024 & 0.0108 & 1.5128 & 0.993136 & -0.963\\
C3 & 2.0 & 300 & 5.2533 & 0.0267 & 5.2821 & 0.994559 & -1.078\\
C3 & 4.0 & 300 & 20.2729 & 0.0892 & 20.3590 & 0.995771 & -0.965\\
C3 & 12.0 & 300 & 180.5886 & 0.7522 & 181.1795 & 0.996739 & -0.785\\
C3 & 40.0 & 300 & 2004.6864 & 8.2815 & 2010.5128 & 0.997102 & -0.704\\
\addlinespace[0.4ex]
\multicolumn{8}{@{}l}{\textbf{Case C4}}\\*
C4 & 0.0 & 300 & 0.1094 & 0.0031 & 0.1136 & 0.962779 & -1.381\\
C4 & 0.5 & 300 & 0.3929 & 0.0040 & 0.3920 & 1.002225 & 0.217\\
C4 & 1.0 & 300 & 1.2351 & 0.0068 & 1.2273 & 1.006365 & 1.149\\
C4 & 2.0 & 300 & 4.5954 & 0.0174 & 4.5682 & 1.005955 & 1.563\\
C4 & 4.0 & 300 & 18.0199 & 0.0586 & 17.9318 & 1.004909 & 1.502\\
C4 & 12.0 & 300 & 161.1026 & 0.4913 & 160.4773 & 1.003897 & 1.273\\
C4 & 40.0 & 300 & 1788.1421 & 5.3833 & 1781.9318 & 1.003485 & 1.154\\
\addlinespace[0.4ex]
\multicolumn{8}{@{}l}{\textbf{Case C5}}\\*
C5 & 0.0 & 300 & 0.6630 & 0.0111 & 0.6897 & 0.961411 & -2.407\\
C5 & 0.5 & 300 & 1.0800 & 0.0144 & 1.1121 & 0.971124 & -2.237\\
C5 & 1.0 & 300 & 2.3438 & 0.0235 & 2.3793 & 0.985060 & -1.515\\
C5 & 2.0 & 300 & 7.4121 & 0.0582 & 7.4483 & 0.995138 & -0.622\\
C5 & 4.0 & 300 & 27.7114 & 0.1954 & 27.7241 & 0.999539 & -0.065\\
C5 & 12.0 & 300 & 244.4116 & 1.6542 & 244.0000 & 1.001687 & 0.249\\
C5 & 40.0 & 300 & 2710.1992 & 18.2392 & 2704.1379 & 1.002241 & 0.332\\
\addlinespace[0.4ex]
\multicolumn{8}{@{}l}{\textbf{Case C6}}\\*
C6 & 0.0 & 300 & 0.2517 & 0.0037 & 0.2532 & 0.994411 & -0.386\\
C6 & 0.5 & 300 & 0.5644 & 0.0046 & 0.5665 & 0.996360 & -0.447\\
C6 & 1.0 & 300 & 1.5020 & 0.0074 & 1.5063 & 0.997121 & -0.588\\
C6 & 2.0 & 300 & 5.2521 & 0.0184 & 5.2658 & 0.997386 & -0.746\\
C6 & 4.0 & 300 & 20.2517 & 0.0627 & 20.3038 & 0.997432 & -0.831\\
C6 & 12.0 & 300 & 180.2429 & 0.5337 & 180.7089 & 0.997422 & -0.873\\*
C6 & 40.0 & 300 & 2000.1224 & 5.8827 & 2005.3165 & 0.997410 & -0.883\\*
\multicolumn{8}{@{}l@{}}{\makebox[0pt][l]{\begin{minipage}[t]{\linewidth}\vspace{0.2em}\hrule height\heavyrulewidth\vspace{0.35em}\footnotesize\textit{Notes.} Ratio is empirical/exact risk; the last column is their difference divided by empirical MCSE. The same SM is used for both FM families.\par\end{minipage}}}\\
\end{longtable}\par
\endgroup

\endgroup

Table~\ref{tab:ultra-hd-lasso-diagnostics} records convergence and support
diagnostics.  All 15,000 LASSO FM fits have zero solver error status.
The largest recorded KKT violation is $1.283\times10^{-6}$, below the frozen $10^{-5}$
threshold, and the largest pass count is 23,910 of the 100,000 allowed.
There are no recorded warnings.  In C5, however, 12 of 300 replications
select the upper LASSO penalty $\lambda=10$ on the independent null-tuning
sample and still return a zero-support FM at $\Delta=40$.  They are valid
converged outcomes and remain in the reported MSE.  Their replication IDs
are 4, 39, 44, 95, 101, 122, 131, 178, 248, 256, 271, and 279.  These 12
losses account for 54.25\% of that cell's total LASSO coefficient loss;
removing them or retuning at the alternative would change the procedure.

\begingroup
\fontsize{9}{11}\selectfont
\setlength{\tabcolsep}{2.3pt}
\setlength{\LTpre}{6pt}
\begingroup
\setlength{\tabcolsep}{2.3pt}
\renewcommand{\arraystretch}{1.10}
\setlength{\LTleft}{0pt plus 1fil}
\setlength{\LTright}{0pt plus 1fil}
\footnotesize

\begin{longtable}{@{\extracolsep{\fill}}lrrrrrrrr@{}}
\caption{\label{tab:ultra-hd-lasso-diagnostics}LASSO FM diagnostics.}\\
\toprule
Case & $\Delta$ & $R$ & Support & Core & Signal & \shortstack[r]{Zero\\fits} & \shortstack[r]{Maximum\\KKT} & \shortstack[r]{Maximum\\passes}\\
\midrule
\endfirsthead
\multicolumn{9}{@{}l}{\textit{\tablename~\thetable. LASSO FM diagnostics. (continued)}}\\
\toprule
Case & $\Delta$ & $R$ & Support & Core & Signal & \shortstack[r]{Zero\\fits} & \shortstack[r]{Maximum\\KKT} & \shortstack[r]{Maximum\\passes}\\
\midrule
\endhead
\midrule
\multicolumn{9}{r}{Continued on next page}\\
\endfoot
\endlastfoot
\multicolumn{9}{@{}l}{\textbf{Case C1}}\\*
C1 & 0.0 & 500 & 85.28 & 6.72 & 0.0140 & 0/500 & 1.24e-06 & 23221\\
C1 & 0.5 & 500 & 85.23 & 6.61 & 0.0460 & 0/500 & 1.18e-06 & 21855\\
C1 & 1.0 & 500 & 85.56 & 6.33 & 0.2560 & 0/500 & 1.27e-06 & 23910\\
C1 & 2.0 & 500 & 84.91 & 5.92 & 0.9220 & 0/500 & 1.26e-06 & 23383\\
C1 & 4.0 & 500 & 82.01 & 5.79 & 1.0000 & 0/500 & 1.28e-06 & 21516\\
C1 & 8.0 & 500 & 73.70 & 5.57 & 1.0000 & 0/500 & 1.17e-06 & 14867\\
C1 & 12.0 & 500 & 65.46 & 5.25 & 1.0000 & 0/500 & 1.19e-06 & 13206\\
C1 & 20.0 & 500 & 50.73 & 4.47 & 1.0000 & 0/500 & 1.17e-06 & 8391\\
C1 & 40.0 & 500 & 29.93 & 2.96 & 1.0000 & 0/500 & 1.10e-06 & 4814\\
\addlinespace[0.4ex]
\multicolumn{9}{@{}l}{\textbf{Case C2}}\\*
C2 & 0.0 & 300 & 86.88 & 15.98 & 0.0533 & 0/300 & 1.19e-06 & 14140\\
C2 & 0.5 & 300 & 87.16 & 15.72 & 0.3567 & 0/300 & 1.18e-06 & 14467\\
C2 & 1.0 & 300 & 87.19 & 15.40 & 0.8200 & 0/300 & 1.28e-06 & 23201\\
C2 & 2.0 & 300 & 86.49 & 15.23 & 1.0000 & 0/300 & 1.20e-06 & 19114\\
C2 & 4.0 & 300 & 84.12 & 15.18 & 1.0000 & 0/300 & 1.07e-06 & 16931\\
C2 & 12.0 & 300 & 70.70 & 14.65 & 1.0000 & 0/300 & 1.18e-06 & 9933\\
C2 & 40.0 & 300 & 33.79 & 10.62 & 1.0000 & 0/300 & 9.19e-07 & 3798\\
\addlinespace[0.4ex]
\multicolumn{9}{@{}l}{\textbf{Case C3}}\\*
C3 & 0.0 & 300 & 72.23 & 4.12 & 0.0000 & 0/300 & 1.09e-06 & 19942\\
C3 & 0.5 & 300 & 72.23 & 3.95 & 0.0133 & 0/300 & 1.11e-06 & 17504\\
C3 & 1.0 & 300 & 72.52 & 3.72 & 0.1367 & 0/300 & 1.04e-06 & 17734\\
C3 & 2.0 & 300 & 72.06 & 3.23 & 0.8133 & 0/300 & 1.09e-06 & 17100\\
C3 & 4.0 & 300 & 66.87 & 3.09 & 1.0000 & 0/300 & 1.09e-06 & 17651\\
C3 & 12.0 & 300 & 41.13 & 2.38 & 1.0000 & 0/300 & 1.10e-06 & 9764\\
C3 & 40.0 & 300 & 12.86 & 0.66 & 1.0000 & 0/300 & 1.09e-06 & 4303\\
\addlinespace[0.4ex]
\multicolumn{9}{@{}l}{\textbf{Case C4}}\\*
C4 & 0.0 & 300 & 82.51 & 8.72 & 0.0133 & 0/300 & 1.07e-06 & 10806\\
C4 & 0.5 & 300 & 83.45 & 8.59 & 0.1300 & 0/300 & 1.20e-06 & 12070\\
C4 & 1.0 & 300 & 84.03 & 8.12 & 0.6033 & 0/300 & 1.15e-06 & 12996\\
C4 & 2.0 & 300 & 83.52 & 7.85 & 0.9733 & 0/300 & 1.12e-06 & 12842\\
C4 & 4.0 & 300 & 80.57 & 7.82 & 1.0000 & 0/300 & 1.14e-06 & 10906\\
C4 & 12.0 & 300 & 63.62 & 7.58 & 1.0000 & 0/300 & 1.01e-06 & 5859\\
C4 & 40.0 & 300 & 23.12 & 5.13 & 1.0000 & 0/300 & 9.53e-07 & 2434\\
\addlinespace[0.4ex]
\multicolumn{9}{@{}l}{\textbf{Case C5}}\\*
C5 & 0.0 & 300 & 52.06 & 3.74 & 0.0067 & 12/300 & 1.05e-06 & 18679\\
C5 & 0.5 & 300 & 52.03 & 3.74 & 0.0433 & 12/300 & 1.10e-06 & 14458\\
C5 & 1.0 & 300 & 52.09 & 3.64 & 0.2133 & 12/300 & 9.10e-07 & 16344\\
C5 & 2.0 & 300 & 50.58 & 3.35 & 0.8367 & 12/300 & 9.32e-07 & 17196\\
C5 & 4.0 & 300 & 43.37 & 3.03 & 0.9600 & 12/300 & 8.94e-07 & 14366\\
C5 & 12.0 & 300 & 16.60 & 1.33 & 0.9600 & 12/300 & 9.02e-07 & 8591\\
C5 & 40.0 & 300 & 4.87 & 0.29 & 0.9600 & 12/300 & 9.24e-07 & 3666\\
\addlinespace[0.4ex]
\multicolumn{9}{@{}l}{\textbf{Case C6}}\\*
C6 & 0.0 & 300 & 171.62 & 13.57 & 0.0033 & 0/300 & 1.21e-06 & 13361\\
C6 & 0.5 & 300 & 171.52 & 13.34 & 0.1333 & 0/300 & 1.11e-06 & 14600\\
C6 & 1.0 & 300 & 171.38 & 12.89 & 0.6600 & 0/300 & 1.13e-06 & 17544\\
C6 & 2.0 & 300 & 169.71 & 12.65 & 1.0000 & 0/300 & 1.15e-06 & 15516\\
C6 & 4.0 & 300 & 163.62 & 12.57 & 1.0000 & 0/300 & 1.08e-06 & 12735\\
C6 & 12.0 & 300 & 128.98 & 11.57 & 1.0000 & 0/300 & 1.05e-06 & 7415\\*
C6 & 40.0 & 300 & 47.57 & 5.65 & 1.0000 & 0/300 & 9.12e-07 & 3608\\*
\multicolumn{9}{@{}l@{}}{\makebox[0pt][l]{\begin{minipage}[t]{\linewidth}\vspace{0.2em}\hrule height\heavyrulewidth\vspace{0.35em}\footnotesize\textit{Notes.} Support and Core are mean selected counts; Signal is the selection proportion for the designated tested coordinate (a null coordinate at $\Delta$ = 0). Zero fits are retained converged all-zero coefficient vectors, not omitted runs. All recorded solver jerr values are zero.\par\end{minipage}}}\\
\end{longtable}\par
\endgroup

\endgroup

Table~\ref{tab:ultra-hd-runtime} reports per-replication stage times from
one TRUBA node with 56 allocated CPUs, 55 concurrent workers, and
single-threaded BLAS/OpenMP.  Of the 2,071 seconds of job wall time,
1,957.875 were spent in scenario wrappers; the remaining 113.125 also
include setup, compilation, finalization, and reporting.  Stage times
measure neither CPU utilization nor the isolated maximum-test cost.

\begingroup
\fontsize{9}{11}\selectfont
\setlength{\tabcolsep}{2.3pt}
\begingroup
\setlength{\tabcolsep}{2.3pt}
\renewcommand{\arraystretch}{1.10}
\setlength{\LTleft}{0pt plus 1fil}
\setlength{\LTright}{0pt plus 1fil}
\footnotesize

\begin{longtable}{@{\extracolsep{\fill}}llrrrrr@{}}
\caption{\label{tab:ultra-hd-runtime}Per-replication computing time.}\\
\toprule
Case & Stage & $R$ & Mean & Median & Q90 & Max\\
\midrule
\endfirsthead
\multicolumn{7}{@{}l}{\textit{\tablename~\thetable. Per-replication computing time. (continued)}}\\
\toprule
Case & Stage & $R$ & Mean & Median & Q90 & Max\\
\midrule
\endhead
\midrule
\multicolumn{7}{r}{Continued on next page}\\
\endfoot
\endlastfoot
\multicolumn{7}{@{}l}{\textbf{Case C1}}\\*
C1 & T1 & 500 & 8.022 & 7.962 & 8.608 & 9.727\\
C1 & T2 & 500 & 19.720 & 19.766 & 20.308 & 20.765\\
C1 & T3 & 500 & 1.497 & 1.405 & 2.117 & 3.336\\
C1 & T4 & 500 & 29.239 & 29.242 & 30.381 & 32.726\\
\addlinespace[0.4ex]
\multicolumn{7}{@{}l}{\textbf{Case C2}}\\*
C2 & T1 & 300 & 1.629 & 1.604 & 1.875 & 2.332\\
C2 & T2 & 300 & 1.261 & 1.175 & 1.422 & 1.828\\
C2 & T3 & 300 & 0.216 & 0.186 & 0.335 & 1.240\\
C2 & T4 & 300 & 3.105 & 3.058 & 3.472 & 3.961\\
\addlinespace[0.4ex]
\multicolumn{7}{@{}l}{\textbf{Case C3}}\\*
C3 & T1 & 300 & 24.427 & 24.985 & 26.174 & 26.885\\
C3 & T2 & 300 & 50.054 & 51.070 & 52.238 & 53.204\\
C3 & T3 & 300 & 2.807 & 2.750 & 3.253 & 5.799\\
C3 & T4 & 300 & 77.288 & 78.936 & 81.047 & 83.378\\
\addlinespace[0.4ex]
\multicolumn{7}{@{}l}{\textbf{Case C4}}\\*
C4 & T1 & 300 & 7.663 & 7.786 & 8.196 & 8.859\\
C4 & T2 & 300 & 15.182 & 15.368 & 15.925 & 16.564\\
C4 & T3 & 300 & 1.078 & 1.078 & 1.280 & 2.074\\
C4 & T4 & 300 & 23.923 & 24.337 & 25.014 & 25.860\\
\addlinespace[0.4ex]
\multicolumn{7}{@{}l}{\textbf{Case C5}}\\*
C5 & T1 & 300 & 8.164 & 8.205 & 8.916 & 9.340\\
C5 & T2 & 300 & 15.579 & 15.752 & 16.255 & 16.692\\
C5 & T3 & 300 & 0.874 & 0.880 & 1.023 & 2.736\\
C5 & T4 & 300 & 24.617 & 24.892 & 25.820 & 27.094\\
\addlinespace[0.4ex]
\multicolumn{7}{@{}l}{\textbf{Case C6}}\\*
C6 & T1 & 300 & 42.641 & 42.769 & 45.804 & 47.811\\
C6 & T2 & 300 & 69.146 & 69.625 & 71.720 & 73.665\\
C6 & T3 & 300 & 3.963 & 3.719 & 5.131 & 8.591\\*
C6 & T4 & 300 & 115.750 & 116.988 & 121.101 & 126.512\\*
\multicolumn{7}{@{}l@{}}{\makebox[0pt][l]{\begin{minipage}[t]{\linewidth}\vspace{0.2em}\hrule height\heavyrulewidth\vspace{0.35em}\footnotesize\textit{Notes.} Times are in seconds, with up to 55 concurrent workers. T1: joint tuning; T2: Ridge FM + SM + maximum test + shrinkage; T3: LASSO FM fitting; T4: recorded compute total. C1 has nine $\Delta$ points, other cases seven. T2 is not isolated test time, and T4 is not job wall time.\par\end{minipage}}}\\
\end{longtable}\par
\endgroup

\endgroup

\clearpage
\section{Operating-characteristic figures}
\label{sec:appendix-operating-figures}

\begin{figure}[htbp]
\centering
\includegraphics[width=\linewidth]{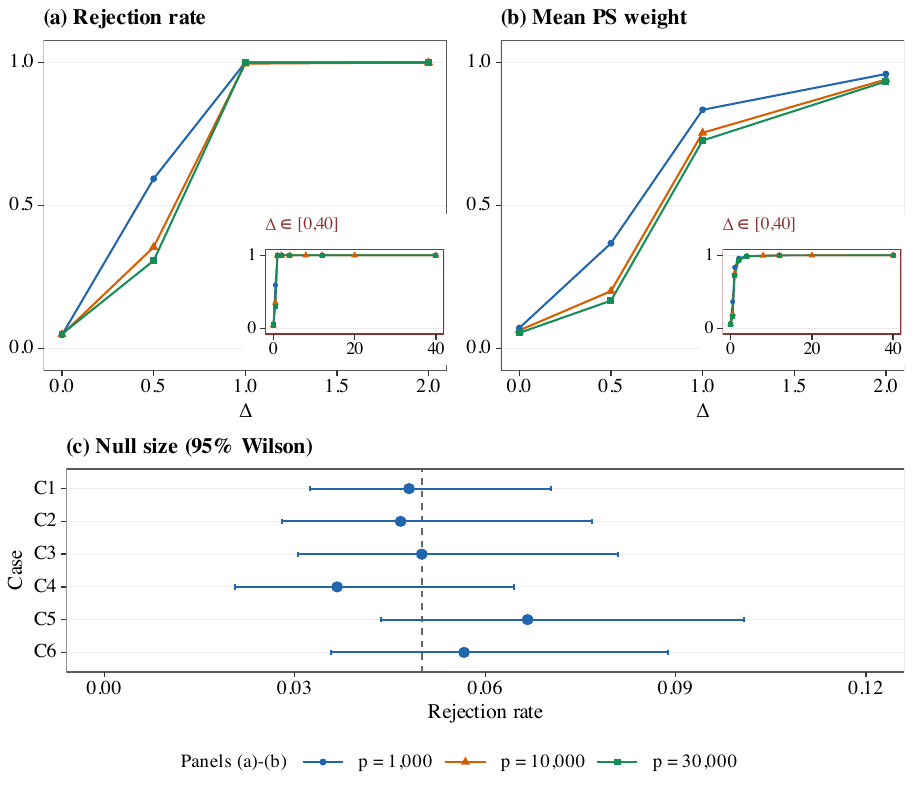}
\caption{Maximum-test operating characteristics.  The first two panels
show rejection rates and mean PS weights for C2, C1, and C3 over
$0\leq\Delta\leq2$; their insets retain the full $0\leq\Delta\leq40$
paths.  The third panel includes null rejection rates for all six cases
with pointwise 95\% Wilson Monte Carlo intervals and nominal level .05.
The departure axis is untransformed.
Rejecting the null does not set the continuous PS weight exactly to one.}
\label{fig:appendix-test-operating}
\end{figure}

Figure~\ref{fig:appendix-test-operating} separates the early power and
weight transitions from the saturated large-departure regime.  At
$\Delta=0.5$, rejection falls from .5933 to .3540 and .3067 as $p$ rises
from 1,000 to 10,000 and 30,000 at fixed $(n,p_1)=(100,20)$.  Thus similar
tails do not imply identical small-departure performance.  Wilson intervals
in the null-size panel describe Monte Carlo uncertainty in empirical
rejection proportions; they are not RPE confidence bands.

\begin{figure}[htbp]
\centering
\includegraphics[width=\linewidth]{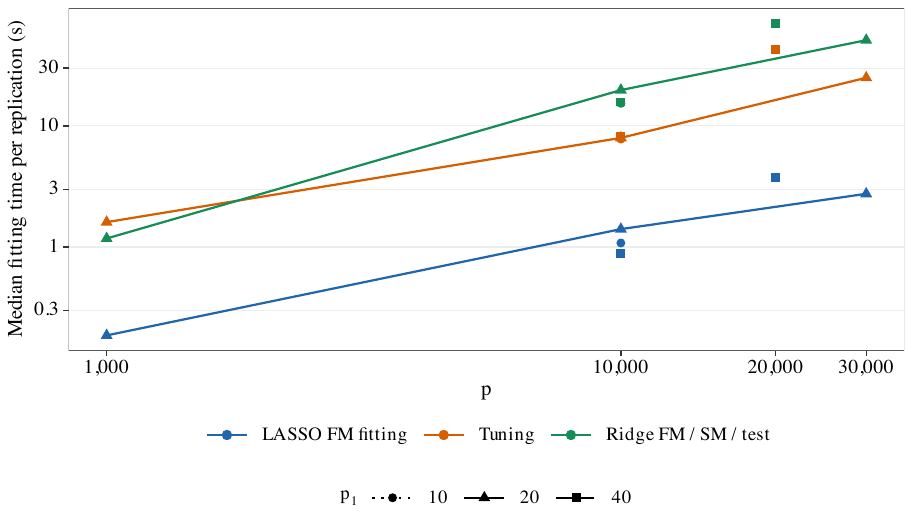}
\caption{Median within-replication fitting times under concurrent execution.
These are not isolated serial benchmarks.  C1 has nine departure points
and C2--C6 seven; the combined stage includes Ridge FM, SM, the test, and
shrinkage.  Axes are logarithmic.}
\label{fig:appendix-runtime}
\end{figure}

Figure~\ref{fig:appendix-runtime} shows the computational cost of increasing
dimension.  The median time for the combined Ridge FM, SM, test, and shrinkage
stage rises from
approximately 1.18 seconds in C2 to 51.07 seconds in C3.  These elapsed
times were measured under concurrent execution and do not isolate the
maximum-test cost.  The C1 path contains nine departures, whereas every
other case contains seven, so not all panels represent equal work per
replication.

\clearpage
\section{DepMap selection and restriction diagnostics}
\label{sec:supplement-depmap}

Table~\ref{tab:depmap-selection} records the four variables retained by
design and distinguishes them from optional CPSS discoveries.  At the frozen
threshold 0.60 both CPSS-LASSO and CPSS-MCP select no optional expression
predictor, so the two selector-labeled SM endpoints coincide.

\begingroup
\setlength{\tabcolsep}{2.3pt}
\renewcommand{\arraystretch}{1.10}
\begin{table}[!htbp]
\small

\caption{\label{tab:depmap-selection}Fixed DepMap submodel features.}
\centering
\begin{tabular}[t]{llll}
\toprule
Feature & Role & CPSS-LASSO & CPSS-MCP\\
\midrule
lineage::Esophagus/Stomach & mandatory & Included & Included\\
lineage::Ovary/Fallopian Tube & mandatory & Included & Included\\
lineage::Uterus & mandatory & Included & Included\\
MSIScore & mandatory & Included & Included\\
\bottomrule
\end{tabular}
\par\vspace{0.35em}
\noindent\begin{minipage}{\linewidth}\footnotesize
\textit{Notes.} Included means retained in the fixed submodel, not discovered by stability selection. Mandatory covariates are specified before CPSS and retained by both selectors. At the 0.60 stability threshold, the optional-extension counts are 0 for CPSS-LASSO and 0 for CPSS-MCP. The mandatory covariates have no CPSS selection frequencies.
\end{minipage}\par
\end{table}
\endgroup

Tables~\ref{tab:depmap-tests} and~\ref{tab:depmap-robust-tests} retain the
full Gaussian and studentized multiplier diagnostics.  The Gaussian law is
the calibration used by the reported PT/S/PS estimators.  The multiplier
calculation is test-only and does not alter their weights.  Its complete
nonrejection, compared with 86--88\% Gaussian rejection across outer fits,
is therefore an assumption-sensitivity result rather than an alternative
post hoc estimator analysis.  In addition, 95 of 100 assessable MCP base
fits lie outside the locally convex path region; MCP remains a sensitivity
selector in this application.

\begingroup
\setlength{\tabcolsep}{2.3pt}
\renewcommand{\arraystretch}{1.10}
\begin{table}[!htbp]
\footnotesize

\caption{\label{tab:depmap-tests}Conditional-Gaussian restriction diagnostics.}
\centering
\begin{tabular}[t]{lrrrrrlrrr}
\toprule
Selector & $p_1$ & $q_{\mathrm{eff}}$ & $T_{\max}$ & $c_{1-\alpha}$ & $p_{\mathrm{MC}}$ & Decision & $\kappa$ & $w_{\mathrm{PS}}$ & \shortstack[r]{MC\\resolution}\\
\midrule
LASSO & 4 & 19107 & 8.4104 & 4.8993 & 0.0010 & Reject & 17.0954 & 0.7583 & 0.0010\\
MCP & 4 & 19107 & 8.4104 & 4.8837 & 0.0010 & Reject & 17.0416 & 0.7591 & 0.0010\\
\bottomrule
\end{tabular}
\par\vspace{0.35em}
\noindent\begin{minipage}{\linewidth}\footnotesize
\textit{Notes.} LASSO and MCP denote the CPSS selectors. Reject/Retain denotes rejection/nonrejection of the restriction at the prespecified nominal level using the finite Monte Carlo p-value. Retain is not acceptance of the null. MC resolution is the p-value grid spacing. Only this conditional-Gaussian test determines the reported shrinkage weights.
\end{minipage}\par
\end{table}
\endgroup

\begingroup
\setlength{\tabcolsep}{2.3pt}
\renewcommand{\arraystretch}{1.10}
\begin{table}[!htbp]
\small

\caption{\label{tab:depmap-robust-tests}Robust test-only sensitivity diagnostics.}
\centering
\begin{tabular}[t]{lrrrlr}
\toprule
Selector & $T_{\max}^{\mathrm{rob}}$ & $c_{1-\alpha}^{\mathrm{rob}}$ & $p_{\mathrm{MC}}^{\mathrm{rob}}$ & Decision & \shortstack[r]{MC\\resolution}\\
\midrule
LASSO & 3.5760 & 4.3805 & 0.4630 & Retain & 0.0002\\
MCP & 3.5760 & 4.4197 & 0.4640 & Retain & 0.0002\\
\bottomrule
\end{tabular}
\par\vspace{0.35em}
\noindent\begin{minipage}{\linewidth}\footnotesize
\textit{Notes.} LASSO and MCP denote the CPSS selectors. Reject/Retain denotes rejection/nonrejection at the prespecified nominal level using the robust multiplier Monte Carlo p-value. MC resolution is the p-value grid spacing. These test-only results never determine shrinkage weights.
\end{minipage}\par
\end{table}
\endgroup

\clearpage
\section*{Reference note}
Bibliographic details for works cited in this supplement are listed in the
combined reference list below.

\bibliographystyle{agsm}
\bibliography{bibliography}
\end{document}